\documentclass[epj,nopacs]{svjour}

\usepackage[10pt]{type1ec}
\usepackage[T1]{fontenc}
\usepackage[utf8]{inputenc}
\usepackage[english]{babel}
\usepackage{amsmath,amssymb}
\usepackage{graphicx}
\usepackage{cite}
\usepackage{microtype}
\usepackage{xurl}
\usepackage[hidelinks,pdfauthor={D. A. Yerokhin},pdftitle={Phantom-Divide Crossing in Barrow-Tsallis Holographic Dark Energy with a Scale-Dependent Barrow Exponent},pdfkeywords={Holographic dark energy, Barrow-Tsallis entropy, Phantom-divide crossing, Future event horizon, Cosmological dynamical systems, Generalized second law}]{hyperref}

\journalname{The European Physical Journal C}

\begin{document}

\title{Phantom-Divide Crossing in Barrow--Tsallis Holographic Dark Energy with a Scale-Dependent Barrow Exponent}
\titlerunning{Phantom-divide crossing in Barrow--Tsallis dark energy}
\author{D.~A.~Yerokhin\thanks{Corresponding author: \email{denyerokhin@gmail.com}}}
\authorrunning{D.~A.~Yerokhin}
\institute{V.~N.~Karazin Kharkiv National University, 4 Svobody Sq., Kharkiv, 61022, Ukraine}
\date{}

\abstract{
Phantom-divide crossing is studied in spatially flat cosmologies with standard Friedmann equations, separately conserved holographic dark energy and a barotropic fluid, and the future event horizon as the infrared cutoff. The horizon entropy and dark-energy density are prescribed by $ S_h=S_{\mathrm{BH}}F(L) $ and $ \rho_X\propto F(L)/L^2 $, where $ F(L) $ is smooth, positive and single-valued, with $ \chi=\mathrm{d}\ln F/\mathrm{d}\ln L $. For $ w_b>-1 $, $ 0<\Omega_X<1 $ and $ \chi<2 $ along the trajectory, at most one phantom-divide crossing is possible. If a crossing occurs, dark energy passes from the quintessence regime into the phantom regime at a maximum of the horizon radius. The crossing kinematics determines the local entropy-scaling dimension and the sign of the local Chevallier-Polarski-Linder coefficient.
Integrating the locally prescribed scale-dependent Barrow exponent reconstructs the Barrow--Tsallis entropy function. A necessary and sufficient event-horizon consistency criterion selects admissible solutions, whose late-time asymptotics include de Sitter states, Type III and Big Rip singularities, and accelerated power-law expansion under the corresponding parameter conditions. A numerical comparison with a constant Barrow exponent distinguishes the occurrence of a crossing from the effect of exponent variation on late-time evolution.
With an adiabatic barotropic fluid, additive total entropy and an equilibrium Gibbs relation for dark energy, the generalized second law fails at the kinematic crossing for positive barotropic-fluid entropy and finite nonzero dark-energy temperature. The divergence of $ c_a^2 $ shows that the purely barotropic adiabatic description of perturbations breaks down. This divergence alone does not establish physical instability.
}

\maketitle
\raggedbottom

\section{Introduction}

The joint analysis of the DESI DR2 baryon acoustic oscillation
(BAO) data together with cosmic microwave background (CMB) and Type Ia
supernova data has renewed interest in dynamical dark energy \cite{DESI2024,DESI2025}. Its
equation of state is characterized by the parameter $w=p/\rho$, where $p$ and
$\rho$ are the pressure and the energy density.
In the Chevallier-Polarski-Linder (CPL) parametrization
\cite{ChevallierPolarski2001,Linder2003}
\begin{equation}
w(a)=w_0+w_a\left(1-\frac{a}{a_0}\right),                       \label{eq:CPL}
\end{equation}
where $a$ is the scale factor and $a_0$ its
present day value.
The coefficient $w_0$ gives the value of $w$ at the present epoch, while $w_a$
characterizes its evolution.
The values $w_0>-1$, $w_a<0$ and $w_0+w_a<-1$ correspond to a crossing from the phantom regime into
the region $w>-1$. In the best-fit CPL models of the DESI DR2 data this crossing
occurs at redshift $z\sim0.4$--$0.5$
\cite{DESI2025,DESIExtended2025}, where $z=a_0/a-1$. A similar
picture persists for several methods of reconstructing $w(z)$; the
exact location of the crossing and its statistical significance -- from
$2.8\sigma$ to $4.2\sigma$ -- nevertheless depend on the supernova compilation,
the parametrization and the choice of priors
\cite{DESI2025,Toomey2026}. The sensitivity to supernova calibration and
selection systematics is discussed separately
\cite{Efstathiou2025,Turyshev2026}.
At the same time, physically motivated models can describe the same data without
requiring a crossing \cite{DindaMaartens2025}. The differing estimates of the
matter density from the CMB and from BAO play an essential
role \cite{ShlivkoPoulin2026}. The reconstructed $ w(z) $ characterizes the background expansion but does not uniquely determine the dark-sector dynamics or require a fundamental phantom field \cite{Mishra2026}. Interactions within the dark sector provide another possible interpretation \cite{LiEtAl2026Interactions}.

For a single local scalar field with Lagrangian density
$p(\phi,X)$, which coincides with the field pressure in a homogeneous and
isotropic Universe, a stable crossing of the phantom divide
is generically impossible \cite{Vikman2005}. Here $\phi$ denotes the scalar field and $X$ its kinetic invariant; the subscript $X$ used below labels the dark-energy component and is unrelated to this invariant. Restrictions on a single local scalar field and possible alternatives with nonlinear kinetic terms are discussed in \cite{NojiriOdintsovOikonomou2026}. Phantom-divide crossing has also been studied in other theories of gravity \cite{NojiriOdintsovOikonomou2025} and in models containing both quintessence and phantom fields \cite{Quintom2010,Quintom2024}. Recent models invoke nonlinear kinetic terms or a field-dependent dark-matter mass \cite{Chen2026} and interactions within the dark sector
\cite{ShahMukherjeePal2025,GuedezounmeDindaMaartens2026};
the same expansion regime has also been obtained in scalar--tensor and vector--tensor theories of gravity \cite{Tsujikawa2026}, as well as in two-field quintom systems confronted with the DESI DR2 data \cite{Thanankullaphong2026}.

Generalized infrared (IR) cutoffs have been considered in \cite{NojiriOdintsov2006}. With the future event horizon as the cutoff, phantom-divide crossing is already possible in standard holographic dark energy \cite{CKN1999,Li2004,Wang2017}. Generalized entropies include the Tsallis and Barrow forms as limiting cases \cite{NojiriOdintsovFaraoni2022,NojiriOdintsovPaul2022GeneralizedEntropy}. Holographic models based on these entropies have been studied for constant values of the Tsallis nonextensivity parameter $ \delta $ and the Barrow exponent $ \Delta $, as well as for scale-dependent extensions \cite{TsallisHDE2018,Nojiri2019,BarrowHDE2020,Basilakos2025}. Interacting Barrow models have also been studied \cite{MamonPaliathanasisSaha2021,GuinPaulGangopadhyay2025}, whereas the present analysis assumes separate conservation of the components. For the Tsallis entropy the conclusions are sensitive to the choice of IR cutoff. Scenarios based on the
Hubble scale or on the Granda-Oliveros cutoff are incompatible with the
standard early-time evolution \cite{IbarboPerlazaEtAl2026}. The variant based on the
future event horizon is, for part of the parameter space, compatible with the
observed growth of structure \cite{Das2026}. In the standard model with the
event-horizon cutoff, combinations of DESI and CMB data
favor a crossing from the quintessence side \cite{LiHDE2025}.
Supernova-dominated data sets, by contrast, correspond to a quintessence
regime without any crossing \cite{Naik2026}. Early- and late-time data
require mutually incompatible evolution regimes for $w(z)$, which limits the
viability of the model \cite{WuHDE2025}. Barrow and Tsallis models with
constant exponents are consistent with the DESI DR2 data but are not favored
over $\Lambda$CDM by information criteria
\cite{LucianoPaliathanasisSaridakis2026}.
A general formalism for holographic dark energy with an arbitrary
entropy function, together with its confrontation with data, is developed in
\cite{Cimdiker2025}, while in \cite{Basilakos2025} the varying Barrow
exponent is prescribed directly as a function of redshift and the
resulting system is studied numerically. Zapata et al. use spline reconstruction for an entropy exponent depending on redshift \cite{ZapataEtAl2026}. Such a dependence on the scale factor need not define a single-valued $ F(L) $ across an extremum of $ L $. Here the entropy function is reconstructed by integrating its local logarithmic slope with respect to the horizon radius. We derive an inverse relation for the local entropy-scaling dimension, a bound on the number of kinematic crossings, and a necessary and sufficient condition for a solution of $ \dot L=HL-1 $ to satisfy the integral definition of the future event horizon.

The analysis assumes standard Friedmann equations, two noninteracting components, the future event horizon as the IR cutoff, and a smooth positive, single-valued entropy factor $ F(L) $, with the dark-energy density prescribed by \eqref{eq:rhoX}. The scale-dependent Barrow exponent provides a specific application. The test of the generalized second law of thermodynamics (GSL) assumes an adiabatic barotropic fluid, additive total entropy and an equilibrium Gibbs relation for dark energy. The perturbation analysis is limited to the purely barotropic adiabatic description. No microscopic model of dark energy is constructed.

\section{Holographic Dark Energy with Generalized Entropy}

To describe the cosmological evolution we adopt the spatially flat
Friedmann--Lema\^{\i}tre-Robertson-Walker metric. Assuming
that the homogeneous and isotropic Universe is filled by two components --
dark energy $X$ and a barotropic fluid -- we arrive at the following
system of equations:
\begin{equation}
3M_{\mathrm{Pl}}^2H^2=\rho_b+\rho_X,
\label{eq:friedmann}
\end{equation}
\begin{equation}
\dot\rho_b+3H(1+w_b)\rho_b=0,\quad w_b=\mathrm{const},
\label{eq:bcont}
\end{equation}
\begin{equation}
\dot\rho_X+3H(1+w_X)\rho_X=0.
\label{eq:xcont}
\end{equation}
Here $a(t)$ is the dimensionless scale factor, $t$ is cosmic
time, a dot denotes the derivative with respect to $t$, and $H=\dot a/a$ is the Hubble
parameter. The reduced Planck mass is defined by
$M_{\mathrm{Pl}}^{-2}=8\pi G$, where $G$ is the gravitational constant.
The quantities $\rho_b$ and $\rho_X$ denote the energy densities of the
barotropic fluid and of dark energy, while $w_b=p_b/\rho_b$ and $w_X=p_X/\rho_X$ are
the corresponding equation-of-state parameters; $p_b$ and $p_X$ denote
the pressures of these components. The subscripts $b$ and $X$ retain this
meaning throughout. We also introduce the variable $N=\ln a$; a prime
denotes the derivative $\mathrm{d}/\mathrm{d}N$.

The absence of energy exchange between dark energy $X$ and the barotropic fluid means that their continuity equations \eqref{eq:bcont}
and \eqref{eq:xcont} hold independently of each other. The values $w_b=0$ and $w_b=1/3$ correspond to a pressureless-matter background and a radiation background, respectively; these two cases are treated separately. In the region $ 0<\Omega_X<1 $, the condition $ w_b>-1 $ ensures that the kinematic level $ HL=1 $ can be crossed at most once. A consistent description of the early-time asymptotics requires $ w_b>-1/3 $.

The IR cutoff is chosen to be the radius of the future event
horizon. This choice ties the local dark-energy density to
the entire subsequent expansion history and therefore requires a future boundary
condition:
\begin{equation}
L(t)=a(t)\int_t^{t_f}\frac{\mathrm{d} t'}{a(t')},
\quad \dot L=HL-1,                                             \label{eq:event-horizon}
\end{equation}
where $t'$ is the integration variable and $t_f$ is the terminal time
of the cosmological evolution under consideration. For a regular eternal future, $t_f=\infty$. The differential equation in \eqref{eq:event-horizon} follows from the integral definition but does not by itself reproduce it. Equivalence is ensured by the condition
\begin{equation}
\lim_{t\to t_f}\frac{L(t)}{a(t)}=0.                             \label{eq:terminal}
\end{equation}
Integrating the differential equation in \eqref{eq:event-horizon}
introduces a free homogeneous term $Ca$ with an arbitrary constant $C$. The boundary condition \eqref{eq:terminal} fixes $C = 0$ and thereby singles out the event horizon uniquely. A similar approach to defining the horizon through a boundary condition was considered in \cite{KimLeeLee2012}. Two circumstances motivate this choice of cutoff. For $F=1$ the Hubble scale $L=H^{-1}$ does not lead to accelerated expansion \cite{Hsu2004}, whereas the event horizon yields realistic late-time dynamics \cite{Li2004}. Moreover, the crossing mechanism developed below relies on the kinematic relation $\dot L=HL-1$ from \eqref{eq:event-horizon}: the branch $HL=1$ passes through an extremum of the event-horizon radius. The nonlocality intrinsic to this choice is later converted into the explicit selection criterion \eqref{eq:global-criterion}.

It is convenient to factor the horizon entropy into its ordinary geometric part and an entropy modification factor:
\begin{equation}
S_h(L)=S_{\mathrm{BH}}(L)F(L),\quad F(L)>0,                   \label{eq:entropy}
\end{equation}
where $S_h$ is the horizon entropy,
$S_{\mathrm{BH}}\propto M_{\mathrm{Pl}}^2L^2$ is the Bekenstein--Hawking
entropy, and $F$ is a dimensionless, sufficiently smooth
function. We prescribe the effect of the generalized entropy on the dark-energy density by
the relation $\rho_XL^4\propto S_h$
\cite{Jahromi2018,NojiriOdintsovPaul2022,Cimdiker2025}. It preserves the
dimensional structure of the original holographic estimate and, for $F=1$,
reduces to the saturated Cohen--Kaplan--Nelson bound. For
an arbitrary function $F$ this relation is an independent model
assumption; the original holographic bound does not fix
it uniquely \cite{CKN1999,Li2004,Wang2017}. Explicitly, the adopted relation reads
\begin{equation}
\rho_X=3c_{H}^2M_{\mathrm{Pl}}^2\frac{F(L)}{L^2}.      \label{eq:rhoX}
\end{equation}
The constant $c_{H}>0$ is dimensionless, and the normalization of $F$ at a single
point can be absorbed into the value of this constant.%
\footnote{Relation \eqref{eq:rhoX} can be compared with the first law of thermodynamics at the horizon: assigning the horizon a temperature $T_h\propto L^{-1}$ and defining the energy as $E=\int_0^{L}T_h\,\mathrm{d}S_h$, one obtains, provided the integral converges at the lower limit, $E(L)\propto M_{\mathrm{Pl}}^2\big[LF(L)+\int_0^{L}F(\ell)\,\mathrm{d}\ell\big]$. For constant $\chi>-1$ this gives $E\propto M_{\mathrm{Pl}}^2(2+\chi)(1+\chi)^{-1}F(L)\,L$, which upon division by the volume reproduces \eqref{eq:rhoX} up to an overall normalization absorbed into $c_{H}$. For varying $\chi$ the thermodynamic expression is a nonlocal functional of $F$ and does not in general coincide with \eqref{eq:rhoX}; in that case \eqref{eq:rhoX} likewise remains an independent model assumption. If in the ultraviolet (UV) limit $F(L)\propto L^{\chi_{\mathrm{UV}}}$ with $\chi_{\mathrm{UV}}\le-1$, the integral requires an ultraviolet cutoff; the condition $\chi_{\mathrm{UV}}>-1$ arises independently below as the condition for a standard matter-dominated era \eqref{eq:matter-early}.}
To describe the local
scale dependence of the entropy we introduce a fixed normalization
scale $L_*>0$ and the quantities
\begin{equation}
\begin{array}{l}
\displaystyle y=\ln\frac{L}{L_*},\quad
\chi(y)=\frac{\mathrm{d}\ln F}{\mathrm{d} y}, \\[2pt]
\displaystyle d_S(y)=2+\chi(y).
\end{array}
\label{eq:definitions}
\end{equation}
The quantity $y$ is the logarithmic horizon-size variable, $\chi$ is the logarithmic
derivative of $F$, and $d_S=\mathrm{d}\ln S_h/\mathrm{d}\ln L$ is the
local entropy-scaling dimension with respect to the radius. The latter is not the spectral
or the Hausdorff dimension of spacetime. It is convenient to write the dynamics
in terms of the dark-energy density parameter $\Omega_X$ and the
dimensionless variable $u$, the inverse of the product $HL$:
\begin{equation}
\Omega_X=\frac{\rho_X}{3M_{\mathrm{Pl}}^2H^2},\quad
u\equiv\frac{1}{HL}=\frac{1}{c_{H}}\sqrt{\frac{\Omega_X}{F(y)}}.       \label{eq:u}
\end{equation}
The kinematic horizon equation fixes the derivative $y'$;
logarithmic differentiation of \eqref{eq:rhoX} gives
$\rho_X'$, and substituting the result into \eqref{eq:xcont} determines
$w_X$:
\begin{equation}
y'=1-u,
\label{eq:yprime}
\end{equation}
\begin{equation}
\frac{\rho_X'}{\rho_X}=(-2+\chi)(1-u),
\label{eq:rhoprime}
\end{equation}
\begin{equation}
w_X=-1+\frac{2-\chi}{3}(1-u).
\label{eq:w-general}
\end{equation}
For $F=1$, that is, $\chi=0$, relation \eqref{eq:w-general} reduces to the familiar expression $w_X=-\tfrac13-\tfrac{2}{3c_{H}}\sqrt{\Omega_X}$ of the standard holographic model \cite{Li2004}.

The separate conservation laws for the two components, together with the Friedmann equation, yield the evolution equation for the dark-energy density parameter $\Omega_X$:
\begin{equation}
\Omega_X'=3(w_b-w_X)\Omega_X(1-\Omega_X),                                  \label{eq:omegaprime}
\end{equation}
and after substituting \eqref{eq:w-general} one obtains a closed au\-ton\-o\-mous
system for the variables $(y,\Omega_X)$:
\begin{equation}
y'=1-\frac{1}{c_{H}}\sqrt{\frac{\Omega_X}{F(y)}},
\label{eq:aut1}
\end{equation}
\begin{equation}
\Omega_X'=\Omega_X(1-\Omega_X)\big[1+3w_b+\chi+(2-\chi)u\big].
\label{eq:aut2}
\end{equation}
Recall that $u$ is defined in \eqref{eq:u}; both right-hand sides depend only on $y$ and $\Omega_X$.
The freedom in the normalization of $F$ noted in the derivation of \eqref{eq:rhoX}
formally means that the background dynamics is invariant under the
rescaling
\begin{equation}
F(L)\longmapsto\lambda F(L),\quad
c_{H}\longmapsto\frac{c_{H}}{\sqrt{\lambda}},\quad \lambda>0,
\label{eq:normalization-symmetry}
\end{equation}
under which neither the product $c_{H}^2F$, nor $\chi$, nor
the density \eqref{eq:rhoX}, nor the right-hand sides of the autonomous system
\eqref{eq:aut1}--\eqref{eq:aut2} change. The expansion history therefore determines
only the product $c_{H}^2F$; the constant $c_{H}$ and the absolute
normalization of $F$ remain separately free, and the condition
$F(L_*)=1$ fixes this ambiguity.
Quantities involving the absolute magnitude of $S_h$ depend on the choice of
normalization.
The system \eqref{eq:aut1}--\eqref{eq:aut2} describes the evolution of the homogeneous Universe.
The equations for dark-energy perturbations require separate assumptions and additional equations.

\section{Phantom-Divide Crossing}

\subsection{Kinematic Branch $HL=1$}

According to \eqref{eq:w-general}, the crossing condition $w_X=-1$ is met
in two cases: for $u=1$ the kinematic factor
$1-u$ vanishes, while for $\chi=2$ the entropic factor $2-\chi$ does. The branch $u=1$
will be called the kinematic branch. For $u=1$, Eq.~\eqref{eq:event-horizon} gives
$\dot L=0$, so on this branch the crossing occurs at an
extremum of the event-horizon radius. The subscript $\mathrm{cr}$
hereafter denotes the value of a quantity at the crossing point.
The crossing direction is determined by the derivative $u'$. Introducing the deceleration parameter
$q=-\ddot a/(aH^2)$, we obtain from $u=(HL)^{-1}$
\begin{equation}
\frac{u'}u
=-\frac{H'}H-\frac{L'}L=(1+q)-(1-u)=q+u,
\label{eq:uprime}
\end{equation}
\begin{equation}
\begin{array}{l}
\displaystyle q=-1-\frac{H'}H \\[2pt]
\displaystyle =\frac12\left[1+3w_b(1-\Omega_X)
+3w_X\Omega_X\right].
\end{array}
\label{eq:qgeneral}
\end{equation}
Consider the interior of the physical phase space,
$0<\Omega_X<1$, which corresponds to the simultaneous presence of both
components. Substituting $u_{\mathrm{cr}}=1$ into
\eqref{eq:uprime}, \eqref{eq:qgeneral} and \eqref{eq:omegaprime} gives
\begin{equation}
\begin{array}{l}
\displaystyle q_{\mathrm{cr}}
=\frac12\left[1+3w_b\right. \\[2pt]
\displaystyle \quad\left.-3(1+w_b)\Omega_{X,\mathrm{cr}}\right],
\end{array}
\label{eq:qcr}
\end{equation}
\begin{equation}
\begin{array}{l}
\displaystyle u'_{\mathrm{cr}}
=q_{\mathrm{cr}}+1 \\[2pt]
\displaystyle =\frac32(1+w_b)(1-\Omega_{X,\mathrm{cr}}) \\[2pt]
\displaystyle \equiv A_{\mathrm{cr}}>0,
\end{array}
\label{eq:ucr}
\end{equation}
\begin{equation}
\begin{array}{l}
\displaystyle \Omega'_{X,\mathrm{cr}}
=3(1+w_b)\Omega_{X,\mathrm{cr}} \\[2pt]
\displaystyle \quad\times(1-\Omega_{X,\mathrm{cr}})>0.
\end{array}
\label{eq:omcr}
\end{equation}
The positivity of $A_{\mathrm{cr}}$ follows from $w_b>-1$ and
$0<\Omega_{X,\mathrm{cr}}<1$. The equality $u_{\mathrm{cr}}=1$ together with
\eqref{eq:u} yields the necessary local condition for the existence of a
crossing:
\begin{equation}
\Omega_{X,\mathrm{cr}}=c_{H}^2F(y_{\mathrm{cr}})<1,   \label{eq:crossing-existence}
\end{equation}
and hence the inequality $c_{H}^2F(y)\geq1$, if it holds for
all $y$, excludes such a crossing in the interior of the physical phase space. For $F=1$, condition \eqref{eq:crossing-existence} reproduces the familiar criterion $c_{H}<1$ for reaching the phantom regime in the standard model \cite{Li2004}.
Accelerated expansion corresponds to $q_{\mathrm{cr}}<0$. In view of
\eqref{eq:qcr} this condition reads
\begin{equation}
\Omega_{X,\mathrm{cr}}>
\frac{1+3w_b}{3(1+w_b)}.                                      \label{eq:acc-condition}
\end{equation}
For a pressureless-matter background ($w_b=0$) the threshold equals $1/3$,
and for radiation ($w_b=1/3$) it equals $1/2$.

To determine the crossing direction we differentiate
\eqref{eq:w-general}:
\begin{equation}
w_X'=-\frac{\chi'}{3}(1-u)-\frac{2-\chi}{3}u'.                 \label{eq:wprime}
\end{equation}
At $u=1$ the term containing $\chi'$ vanishes. Substituting
\eqref{eq:ucr} leads to the relation
\begin{equation}
w'_{X,\mathrm{cr}}=-\frac{2-\chi_{\mathrm{cr}}}{2}
(1+w_b)(1-\Omega_{X,\mathrm{cr}}).                                 \label{eq:orientation}
\end{equation}
For $w_b>-1$ and $\chi_{\mathrm{cr}}<2$ the derivative is negative, and as
$N$ increases the trajectory crosses from the quintessence regime, $w_X>-1$, into
the phantom regime, $w_X<-1$. The condition $\chi_{\mathrm{cr}}>2$ gives a
phantom-to-quintessence crossing.\footnote{Relations \eqref{eq:ucr} and \eqref{eq:orientation} generalize to noninteracting barotropic fluids with constant $ w_i>-1 $. At $ u=1 $ one has $ u'_{\mathrm{cr}}=\tfrac32\sum_i(1+w_i)\Omega_{i,\mathrm{cr}}>0 $ and $ w'_{X,\mathrm{cr}}=-\tfrac12(2-\chi_{\mathrm{cr}})\sum_i(1+w_i)\Omega_{i,\mathrm{cr}} $. The kinematic level can therefore be crossed at most once. These local crossing results also apply to a mixture of matter and radiation. The early-time asymptotics are treated separately for each dominant component.}
For the estimates that follow we adopt the range
\begin{equation}
-2<\chi\le1,                                                    \label{eq:corridor}
\end{equation}
where the lower bound ensures that $S_h$ increases with the radius $L$, while the upper one, $d_S\le3$,
coincides with the Barrow limit of maximal fractal deformation of the horizon, in which the entropy grows no faster than the volume \cite{Barrow2020}. Thermodynamic
consistency is checked independently against the GSL
\cite{Bekenstein1974,Davies1987}.

By \eqref{eq:uprime}, at the crossing
$u'_{\mathrm{cr}}=1+q_{\mathrm{cr}}$, and by
\eqref{eq:crossing-existence}
$\Omega_{X,\mathrm{cr}}=c_{H}^2F(y_{\mathrm{cr}})$.
Relations \eqref{eq:qcr} and \eqref{eq:orientation} express the deceleration
parameter $q_{\mathrm{cr}}$ and the derivative $w'_{X,\mathrm{cr}}$ in terms of the
pair $(\Omega_{X,\mathrm{cr}},\chi_{\mathrm{cr}})$; for $w_b>-1$ they
can be solved uniquely for this pair. For a specified $ w_b $, the kinematic quantities $ q_{\mathrm{cr}} $ and $ w'_{X,\mathrm{cr}} $ therefore determine the product $ c_H^2F(y_{\mathrm{cr}}) $ and the local entropy-scaling dimension $ d_{S,\mathrm{cr}} $:
\begin{equation}
\begin{array}{l}
\displaystyle c_{H}^2F(y_{\mathrm{cr}})=1-\frac{2(1+q_{\mathrm{cr}})}{3(1+w_b)}, \\[2pt]
\displaystyle d_{S,\mathrm{cr}}=4+\frac{3w'_{X,\mathrm{cr}}}{1+q_{\mathrm{cr}}}.
\end{array}
\label{eq:crossing-inverse-map}
\end{equation}
We rewrite the second line of \eqref{eq:crossing-inverse-map} in terms of the
derivative with respect to redshift,
\begin{equation}
d_{S,\mathrm{cr}}=4-\frac{3(1+z_{\mathrm{cr}})}{1+q_{\mathrm{cr}}}
\left.\frac{\mathrm{d}w_X}{\mathrm{d}z}\right|_{\mathrm{cr}}.
\label{eq:crossing-redshift-map}
\end{equation}
The right-hand sides of \eqref{eq:crossing-inverse-map} and
\eqref{eq:crossing-redshift-map} contain only kinematic
quantities and the background parameter $w_b$; they depend neither on the form of the function $F$
nor on the choice of parametrization of $w(a)$. If the reconstructed $ d_{S,\mathrm{cr}} $ lies outside $ 0<d_{S,\mathrm{cr}}\le3 $, the range corresponding to \eqref{eq:corridor}, the associated entropy class is excluded within the assumed standard Friedmann dynamics, separate conservation and future-event-horizon cutoff.

Relation \eqref{eq:ucr} bounds the number of crossings on the kinematic branch. For $ w_b>-1 $, along any continuously differentiable trajectory in the region $ 0<\Omega_X<1 $, one has $ u'>0 $ at every point of the level set $ u=1 $. A return from the region $ u>1 $ to the region $ u<1 $ would require a crossing point with $ u'<0 $ or a tangency with $ u'=0 $, which contradicts \eqref{eq:ucr}. The level set $ u=1 $ can therefore be crossed at most once. The absence of a second crossing makes this level set an observational criterion. For solutions with the early-time asymptotics $u\to0$ (see below), the kinematic crossing has already occurred by a given epoch if and only if $\Omega_X>c_{H}^2F(y)$, that is, $u>1$; for $u<1$ it has not yet occurred. The subsequent trajectory may reach the level set $u=1$, approach it asymptotically, or never reach it; the specific late-time outcome is determined by the function $F$ and by the global condition \eqref{eq:global-criterion}
formulated in Sec.~\ref{sec:horizon-criterion}.
The entropic branch $\chi=2$ has a different local structure and is considered below.

\subsection{Local Behavior near the Crossing}

A local expansion establishes the regularity of the transition and
determines the nature of the extrema of the physical quantities. Set
$\tau=N-N_{\mathrm{cr}}$.  Using
$A_{\mathrm{cr}}=\frac32(1+w_b)(1-\Omega_{X,\mathrm{cr}})$, we obtain
\begin{equation}
u=1+A_{\mathrm{cr}}\tau+O(\tau^2),
\label{eq:u-series}
\end{equation}
\begin{equation}
y=y_{\mathrm{cr}}-\frac{A_{\mathrm{cr}}}{2}\tau^2+O(\tau^3),
\label{eq:y-series}
\end{equation}
\begin{equation}
\frac{L}{L_{\mathrm{cr}}}=1-\frac{A_{\mathrm{cr}}}{2}\tau^2+O(\tau^3),
\label{eq:L-series}
\end{equation}
\begin{equation}
w_X+1=-\frac{(2-\chi_{\mathrm{cr}})A_{\mathrm{cr}}}{3}\tau+O(\tau^2),
\label{eq:w-series}
\end{equation}
\begin{equation}
\begin{array}{l}
\displaystyle \Omega_X=\Omega_{X,\mathrm{cr}}+3(1+w_b)\Omega_{X,\mathrm{cr}}
(1-\Omega_{X,\mathrm{cr}})\tau \\[2pt]
\displaystyle \hspace{2.6em}+O(\tau^2).
\end{array}
\label{eq:Om-series}
\end{equation}
The expansions \eqref{eq:y-series} and \eqref{eq:L-series} show that
the horizon radius has a strict local maximum. Substituting the same
expansion into \eqref{eq:rhoX} and \eqref{eq:entropy} gives
\begin{equation}
\frac{\rho_X}{\rho_{X,\mathrm{cr}}}=1+
\frac{(2-\chi_{\mathrm{cr}})A_{\mathrm{cr}}}{2}\tau^2+O(\tau^3),
\label{eq:rho-series}
\end{equation}
\begin{equation}
\frac{S_h}{S_{h,\mathrm{cr}}}=1-
\frac{(2+\chi_{\mathrm{cr}})A_{\mathrm{cr}}}{2}\tau^2+O(\tau^3).
\label{eq:S-series}
\end{equation}
For $-2<\chi_{\mathrm{cr}}<2$ the horizon entropy also has a local
maximum, while the dark-energy density has a local minimum.

\subsection{Entropic Branch and the Degenerate Case}

Equation \eqref{eq:w-general} implies a second way of reaching
$w_X=-1$: the equality $\chi=2$ with $u\ne1$. The notation
$\chi_{,y}=\mathrm{d}\chi/\mathrm{d}y$ is used for the derivative with respect to the
logarithmic horizon-size variable. Since $\chi'=\chi_{,y}(1-u)$, Eq.~\eqref{eq:wprime}
on this branch takes the form
\begin{equation}
\left.w_X'\right|_{\chi=2}=-\frac13\chi_{,y}(1-u)^2.            \label{eq:chi2}
\end{equation}
For $\chi_{,y}<0$ the parameter $w_X$ increases through the value $-1$, that
is, the trajectory undergoes a phantom-to-quintessence crossing.
For $\chi_{,y}>0$ a quintessence-to-phantom crossing occurs. If $\chi_{,y}=0$,
the direction is determined by higher-order terms. The branch
$\chi=2$ lies outside the range \eqref{eq:corridor}.

When $u=1$ and $\chi=2$ hold simultaneously, the linear
term in the expansion of $w_X+1$ vanishes. With
$\chi=2+\beta_{\mathrm{cr}}(y-y_{\mathrm{cr}})+\cdots$, where
$\beta_{\mathrm{cr}}=\chi_{,y}(y_{\mathrm{cr}})$, we have
\begin{equation}
w_X+1=-\frac{\beta_{\mathrm{cr}}A_{\mathrm{cr}}^{2}}{6}\tau^3+O(\tau^4).     \label{eq:cubic}
\end{equation}
The crossing is cubic in $\tau$. For $\beta_{\mathrm{cr}}<0$
a phantom-to-quintessence crossing takes place. For
$\beta_{\mathrm{cr}}>0$ the trajectory undergoes a quintessence-to-phantom
crossing.

\section{Scale Dependence of the Barrow Exponent in the Barrow--Tsallis Entropy}

For constant parameters the composite Barrow--Tsallis entropy ansatz
$S_{\mathrm{BT}}$ \cite{TsallisCirto2013,Barrow2020,TsallisHDE2018,BarrowHDE2020,Bolotin2026}
has a power-law dependence
\begin{equation}
S_{\mathrm{BT}}\propto L^{\delta(2+\Delta)},\quad
\chi=2\delta-2+\delta\Delta.                                   \label{eq:BTconstant}
\end{equation}
Here $\delta$ is the Tsallis nonextensivity parameter and $\Delta$ is the Barrow
exponent.
In these variables condition \eqref{eq:corridor} becomes
$d_S=\delta(2+\Delta)\le3$ and imposes an additional restriction that does not
follow from the original Barrow range
$0\le\Delta\le1$ alone. A simple replacement $\Delta\mapsto\Delta(L)$ \cite{Nojiri2019,Basilakos2025} in the expression
$S\propto L^{\delta[2+\Delta(L)]}$ produces, upon logarithmic
differentiation, an extra term
$\delta\ln(L/\ell_{\mathrm{P}})\,\mathrm{d}\Delta/\mathrm{d}\ln L$,
where $\ell_{\mathrm{P}}$ is the Planck length. As a result, the prescribed
function $\Delta(L)$ no longer coincides with the local power-law
exponent. To preserve the local meaning of the Barrow exponent,
we prescribe the logarithmic derivative of the entropy:
\begin{equation}
\frac{\mathrm{d}\ln S_h}{\mathrm{d}\ln L}=2\delta+
\delta\Delta_{\mathrm{loc}}(L),\quad
\chi=2\delta-2+\delta\Delta_{\mathrm{loc}}.                   \label{eq:localBT}
\end{equation}
The function $\Delta_{\mathrm{loc}}(L)$ denotes the local value of the
Barrow exponent.
Equation \eqref{eq:localBT} can be integrated explicitly:
\begin{equation}
\begin{array}{l}
\displaystyle F(L)=F(L_*) \\[2pt]
\displaystyle \times\exp\left[\int_{L_*}^{L}
\big(2\delta-2+\delta\Delta_{\mathrm{loc}}(\ell)\big)
\frac{\mathrm{d}\ell}{\ell}\right].
\end{array}
\label{eq:F-reconstruction}
\end{equation}
Once a normalization condition is chosen, for instance $F(L_*)=1$,
an admissible profile $\Delta_{\mathrm{loc}}(L)$ reconstructs the
global entropy function uniquely.

\begin{sloppypar}
Constructing an explicit model requires a smooth bound\-ed function
interpolating between two power-law regimes. We assume $\delta>0$ and
$\Delta_{\mathrm{UV}}>0$; if $\Delta_{\mathrm{UV}}$ is interpreted as the original Barrow exponent, the additional requirement $\Delta_{\mathrm{UV}}\le1$ applies. We choose the simplest monotonic profile
\end{sloppypar}
\begin{equation}
\Delta_{\mathrm{loc}}(L)=\frac{\Delta_{\mathrm{UV}}}
{1+(L/L_t)^\kappa},\quad \kappa>0.                             \label{eq:flow}
\end{equation}
The parameter $\Delta_{\mathrm{UV}}$ sets the ultraviolet limit,
$L_t>0$ is the transition scale, and $\kappa$ is the sharpness of the transition. The function
\eqref{eq:flow} tends to $\Delta_{\mathrm{UV}}$ for $L\ll L_t$ and to
zero for $L\gg L_t$. The decreasing profile is consistent with the physical meaning of the Barrow exponent: it encodes the quantum-gravitational fractal deformation of the horizon, which is maximal on small scales and smooths out in the infrared limit \cite{Barrow2020,Basilakos2025}. The form \eqref{eq:flow} is also motivated by an analogy with threshold
decoupling \cite{AppelquistCarazzone1975,Berges2002}; no derivation from a
renormalization-group beta function is implied here. Introducing the
dimensionless variable $x=(L/L_t)^\kappa$, we obtain
\begin{equation}
\chi(L)=2\delta-2+\frac{\delta\Delta_{\mathrm{UV}}}{1+x},
\label{eq:chi-flow}
\end{equation}
\begin{equation}
\frac{\mathrm{d}\chi}{\mathrm{d} y}=-\kappa\delta\Delta_{\mathrm{UV}}
\frac{x}{(1+x)^2}.
\label{eq:chi-derivative}
\end{equation}
Integration with the normalization condition $F(L_*)=1$ gives
\begin{equation}
F(L)=\left(\frac{L}{L_*}\right)^{\chi_{\mathrm{UV}}}
\left[
\frac{1+(L/L_t)^\kappa}{1+(L_*/L_t)^\kappa}
\right]^{-\delta\Delta_{\mathrm{UV}}/\kappa},                         \label{eq:Fglobal}
\end{equation}
where
\begin{equation}
\chi_{\mathrm{UV}}=2\delta-2+\delta\Delta_{\mathrm{UV}},\quad
\chi_{\mathrm{IR}}=2\delta-2.                                         \label{eq:limits}
\end{equation}
The subscripts $\mathrm{UV}$ and $\mathrm{IR}$ denote the small- and
large-horizon-radius limits, respectively. The resulting function $F$ is positive
for all $L>0$ and has the prescribed power-law asymptotics; its boundedness
is not required for the analysis that follows.

For comparison with the local analysis we expand $\chi$ about
the normalization scale $L_0=L_*$. With
$x_0=(L_0/L_t)^\kappa$, the Taylor coefficients are given by
\begin{equation}
\chi(y)=\alpha+\beta y+\frac{\gamma}{2}y^2+O(y^3),
\label{eq:local-series}
\end{equation}
\begin{equation}
\alpha=2\delta-2+\frac{\delta\Delta_{\mathrm{UV}}}{1+x_0},
\label{eq:alpha}
\end{equation}
\begin{equation}
\beta=-\kappa\delta\Delta_{\mathrm{UV}}\frac{x_0}{(1+x_0)^2},
\label{eq:beta}
\end{equation}
\begin{equation}
\gamma=-\kappa^2\delta\Delta_{\mathrm{UV}}
\frac{x_0(1-x_0)}{(1+x_0)^3},
\label{eq:gamma}
\end{equation}
\begin{equation}
\ln F(y)=\alpha y+\frac{\beta}{2}y^2+
\frac{\gamma}{6}y^3+O(y^4).
\label{eq:lnFseries}
\end{equation}
A particularly simple closed form arises for $\kappa=2$ and
$L_t=L_*=L_0$:
\begin{equation}
\chi=\alpha+\beta\tanh y,\quad
F=\mathrm{e}^{\alpha y}(\cosh y)^\beta,                               \label{eq:tanh}
\end{equation}
\begin{equation}
\alpha=2\delta-2+\frac{\delta\Delta_{\mathrm{UV}}}{2},\quad
\beta=-\frac{\delta\Delta_{\mathrm{UV}}}{2}.                         \label{eq:map}
\end{equation}
The parameters $\alpha$ and $\beta$ in \eqref{eq:tanh} coincide with the coefficients of the local expansion \eqref{eq:local-series}.

\subsection{Matter- and Radiation-Era Asymptotics}

The scale deformation is admissible only if the standard
early-time regime, in which the dark-energy contribution is small, is preserved. During an era
dominated by a single barotropic fluid with constant $w_b>-1/3$, the integral in
\eqref{eq:event-horizon} has a finite limit as $a\to0$ provided the
global event-horizon condition holds. The ratio
$L/a$ then tends to a finite positive constant and $L\propto a$.
This yields the asymptotics
\begin{equation}
\rho_X\propto a^{-2+\chi_{\mathrm{UV}}},
\label{eq:rhoearly}
\end{equation}
\begin{equation}
\Omega_X\propto a^{1+3w_b+\chi_{\mathrm{UV}}}
=a^{d_{S,\mathrm{UV}}-1+3w_b},
\label{eq:Omegaearly}
\end{equation}
\begin{equation}
u\propto a^{(1+3w_b)/2}.
\label{eq:uearly}
\end{equation}
The dark-energy contribution becomes negligible toward the past when
\begin{equation}
1+3w_b+\chi_{\mathrm{UV}}>0
\quad\Longleftrightarrow\quad d_{S,\mathrm{UV}}>1-3w_b.               \label{eq:past-condition}
\end{equation}
Here $d_{S,\mathrm{UV}}=2+\chi_{\mathrm{UV}}$ is the ultraviolet
limit of the local entropy-scaling dimension. For the matter- and
radiation-dominated eras the condition reads
\begin{equation}
w_b=0:\quad \Omega_X\propto a^{1+\chi_{\mathrm{UV}}},\quad
\chi_{\mathrm{UV}}>-1,
\label{eq:matter-early}
\end{equation}
\begin{equation}
w_b=\frac13:\quad \Omega_X\propto a^{2+\chi_{\mathrm{UV}}},\quad
\chi_{\mathrm{UV}}>-2.
\label{eq:radiation-early}
\end{equation}
When \eqref{eq:past-condition} holds, both
$\Omega_X\to0$ and $u\to0$. The solution thus reproduces the standard
barotropic asymptotic state of the early Universe. We emphasize that the known early-time constraints on the Barrow exponent, for example $\Delta\lesssim10^{-4}$ from Big Bang nucleosynthesis \cite{BarrowBBN2021}, were obtained for modified Friedmann equations within apparent-horizon thermodynamics.
In the present model the dark-energy contribution to the early-time dynamics is asymptotically suppressed by the power law \eqref{eq:Omegaearly}, so constraints of this kind -- derived for modified Friedmann equations both from Big Bang nucleosynthesis \cite{BarrowBBN2021,SheykhiShahbazi2025} and from joint analyses of late-time observational data \cite{LeonEtAl2021} -- do not carry over directly. Notably, these estimates themselves differ by two orders of magnitude ($\Delta\lesssim10^{-4}$ versus $\Delta\simeq10^{-2}$) depending on how the modified Friedmann equations are derived. Whether values of $\Delta_{\mathrm{UV}}$ of order unity are compatible with nucleosynthesis and recombination requires a separate quantitative check of $\Omega_X$ at the corresponding epochs.

\section{Selection of Physically Admissible Solutions and Late-Time Asymptotic States}

\subsection{Event-Horizon Consistency Criterion}\label{sec:horizon-criterion}

Solutions of the autonomous system generically contain the homogeneous mode $Ca$. To remove it we rewrite the boundary condition \eqref{eq:terminal} in terms of the dynamical variables. From $y'=1-u$
and $N=\ln a$ it follows that
\begin{equation}
\left(\ln\frac{L}{a}\right)'=y'-1=-u.                          \label{eq:La-prime}
\end{equation}
Let $N_i$ be an arbitrary instant, $L_i=L(N_i)$ and
$a_i=a(N_i)$, and let $\widetilde N$ be the integration variable. Integrating
between $N_i$ and $N$ then gives
\begin{equation}
\frac{L(N)}{a(N)}=\frac{L_i}{a_i}
\exp\left[-\int_{N_i}^{N}u(\widetilde N)\,\mathrm{d}\widetilde N\right].
                                                                        \label{eq:La-integral}
\end{equation}
Denote by $N_f$ the value of $N$ as $t\to t_f$. For a positive
finite ratio $L_i/a_i$, condition \eqref{eq:terminal} is equivalent to
\begin{equation}
\int_{N_i}^{N_f}u(N)\,\mathrm{d} N=\infty.                         \label{eq:global-criterion}
\end{equation}
This is a necessary and sufficient condition for a solution of the local
equation $\dot L=HL-1$ to satisfy the integral definition of the
future event horizon as well. The homogeneous mode $Ca$ is thereby removed. The dependence on the
future evolution, intrinsic to the choice of the event-horizon cutoff,
persists.

Consider first the branch $y\to+\infty$ with
$\chi_{\mathrm{IR}}>0$ and $\Omega_X\to1$. Let
$F_{\mathrm{IR}}>0$ denote the coefficient of the large-radius asymptotics. Then
\begin{equation}
F\sim F_{\mathrm{IR}}\mathrm{e}^{\chi_{\mathrm{IR}}y},\quad
u\sim \frac{\mathrm{e}^{-\chi_{\mathrm{IR}}y/2}}{c_{H}\sqrt{F_{\mathrm{IR}}}}.           \label{eq:right-runaway}
\end{equation}
Equation \eqref{eq:right-runaway} gives $y'=1+o(1)$, where $o(1)$ denotes a quantity vanishing in the limit under consideration; hence $y\sim N$ and the integral $\int u\,\mathrm{d}N$ converges. Criterion \eqref{eq:global-criterion} rejects this branch: it
describes the homogeneous mode $Ca$ of the differential equation \eqref{eq:event-horizon}, which is discarded by the boundary condition \eqref{eq:terminal}.

Another limiting regime arises for
$\chi_{\mathrm{UV}}>0>\chi_{\mathrm{IR}}$, that is, when $F$ first
increases and then decreases. In the absence of de Sitter fixed points, $y$ decreases without bound along the physically admissible trajectory ($y\to-\infty$). If $0<\chi_{\mathrm{UV}}<2$ and $\Omega_X\to1$, then
\begin{equation}
\begin{array}{l}
\displaystyle F\sim F_{\mathrm{UV}}\left(\frac{L}{L_*}\right)^{\chi_{\mathrm{UV}}}, \\[2pt]
\displaystyle \rho_X\sim \rho_*\left(\frac{L}{L_*}\right)^{\chi_{\mathrm{UV}}-2},\quad
u\to\infty.
\end{array}
\label{eq:left-asymptotic}
\end{equation}
Here $F_{\mathrm{UV}} > 0$ and $\rho_* > 0$ are constant coefficients of the ultraviolet asymptotics. Since $HL=u^{-1}$, the limiting relations take the form
\begin{equation}
\begin{array}{l}
\displaystyle \dot L=HL-1=u^{-1}-1=-1+o(1), \\[2pt]
\displaystyle L=t_f-t+o(t_f-t), \\[2pt]
\displaystyle H\sim H_*(t_f-t)^{-1+\chi_{\mathrm{UV}}/2}, \\[2pt]
\displaystyle \ln\frac{a_f}{a(t)}
=\int_t^{t_f}H\,\mathrm{d} t
\sim\frac{2H_*}{\chi_{\mathrm{UV}}}(t_f-t)^{\chi_{\mathrm{UV}}/2}.
\end{array}
\label{eq:typeIII-check}
\end{equation}
The constant $H_*>0$ is the coefficient in the asymptotics of the Hubble parameter, and
$a_f$ denotes the finite limiting value of the scale factor.
Since $0<\chi_{\mathrm{UV}}<2$, the parameter $H$ diverges as
$t\to t_f$, while its time integral remains finite. Therefore
\begin{equation}
a(t)\to a_f<\infty,\quad
\rho_X\to\infty,\quad |p_X|\to\infty.                         \label{eq:typeIII}
\end{equation}
This behavior corresponds to a Type III finite-time future singularity in the
standard classification \cite{NojiriOdintsovTsujikawa2005}. The global condition
is also satisfied: $L/a\to0$ and
$\int u\,\mathrm{d} N=\int\mathrm{d} t/L\sim-\ln(t_f-t)\to\infty$.

The boundary value $\chi_{\mathrm{UV}}=0$ forms a separate
asymptotic class. Let $F\to F_{\mathrm{UV}}>0$ as
$y\to-\infty$, $\Omega_X\to1$ and $C_{\mathrm{UV}}\equiv
c_{H}\sqrt{F_{\mathrm{UV}}}<1$; for the profile \eqref{eq:flow} with
$\chi_{\mathrm{UV}}=0$ the function $F$ decreases monotonically, so there are no finite
roots. Then $u\to C_{\mathrm{UV}}^{-1}>1$ and
\begin{equation}
\begin{array}{l}
\displaystyle L=(1-C_{\mathrm{UV}})(t_f-t)+o(t_f-t), \\[2pt]
\displaystyle H\sim\frac{C_{\mathrm{UV}}}{1-C_{\mathrm{UV}}}\,
\frac{1}{t_f-t}, \\[2pt]
\displaystyle a\propto(t_f-t)^{-C_{\mathrm{UV}}/(1-C_{\mathrm{UV}})}, \\[2pt]
\displaystyle w_X\to-1-\frac{2(1-C_{\mathrm{UV}})}{3C_{\mathrm{UV}}}.
\end{array}
\label{eq:typeI-boundary}
\end{equation}
The scale factor, $\rho_X$ and $|p_X|$ diverge at a finite
$t_f$, which corresponds to a Type I (Big Rip) singularity; the global condition
is satisfied because $L\propto t_f-t$ and
$\int u\,\mathrm{d}N=\int\mathrm{d}t/L\to\infty$. Thus
$\chi_{\mathrm{UV}}=0$ separates the Big Rip from the regime
\eqref{eq:typeIII} that arises for $0<\chi_{\mathrm{UV}}<2$.

\subsection{Finite and Non-Hyperbolic Fixed Points}

The regular late-time evolution is governed by the finite fixed points
of the system \eqref{eq:aut1}-\eqref{eq:aut2}. The subscript $\mathrm{dS}$ hereafter denotes the value of a quantity at such a point. The fixed-point conditions read
\begin{equation}
\Omega_{X,\mathrm{dS}}=1,\quad u_{\mathrm{dS}}=1,\quad F(y_{\mathrm{dS}})=c_{H}^{-2}.        \label{eq:dscondition}
\end{equation}
At this point $w_X=-1$ and $H$ is constant, so it describes a
de Sitter state. To determine its stability we linearize the
system in the variables $(y,\Omega_X)$. The Jacobian matrix $J_{\mathrm{dS}}$ and its
eigenvalues $\lambda_1$, $\lambda_2$ are
\begin{equation}
\begin{array}{l}
\displaystyle J_{\mathrm{dS}}=\begin{pmatrix}
\chi_{\mathrm{dS}}/2 & -1/2\\
0 & -3(1+w_b)
\end{pmatrix}, \\[2pt]
\displaystyle \lambda_1=\frac{\chi_{\mathrm{dS}}}{2},\quad
\lambda_2=-3(1+w_b).
\end{array}
\label{eq:eigenvalues}
\end{equation}
For $w_b>-1$ the second eigenvalue is negative. A root on the
decreasing branch of $F$, for which $\chi_{\mathrm{dS}}<0$, is an attractor;
a root on the increasing branch, with $\chi_{\mathrm{dS}}>0$, is a saddle point.
The off-diagonal element of $J_{\mathrm{dS}}$ mixes the two linear modes.
Hence, for $\chi_{\mathrm{dS}}/2\ne-3(1+w_b)$, the asymptotics reads
\begin{equation}
\begin{array}{l}
\displaystyle 1-\Omega_X=C_{\Omega}\mathrm{e}^{-3(1+w_b)N}+\cdots, \\[2pt]
\displaystyle y-y_{\mathrm{dS}}=C_y\mathrm{e}^{\chi_{\mathrm{dS}}N/2}
+C_b\mathrm{e}^{-3(1+w_b)N}+\cdots.
\end{array}
\label{eq:dsasymptotic}
\end{equation}
The constants $C_{\Omega}$, $C_y$ and $C_b$ are fixed by the initial
conditions. In the resonant case $\chi_{\mathrm{dS}}/2=-3(1+w_b)$ the second term in
the expansion of $y-y_{\mathrm{dS}}$ is replaced by a quantity proportional to
$N\mathrm{e}^{-3(1+w_b)N}$.
On the attractor branch the deviation from $w_X=-1$ into the phantom regime decays asymptotically: after the crossing $1-u<0$ and, by \eqref{eq:w-general}, $w_X\to-1^-$, so that the solution is asymptotically de Sitter and no Type I (Big Rip) singularity occurs. By contrast, for the profile \eqref{eq:Fglobal}, in the absence of a finite attractor, with $0<\chi_{\mathrm{UV}}<2$ and $\chi_{\mathrm{IR}}<0$, the regime \eqref{eq:typeIII} is realized.

For the function \eqref{eq:flow} the number of fixed
points is determined by the shape of $F$ and by the value of $c_{H}^{-2}$. For
$\Delta_{\mathrm{UV}}>0$ the possible regimes are listed in
Table~\ref{tab:roots}.

\begin{table*}[t]
\centering
\caption{Number and stability of the roots of $F(y)=c_{H}^{-2}$ for
$\delta>0$ and $\Delta_{\mathrm{UV}}>0$.}
\label{tab:roots}
\begin{tabular}{|p{0.25\textwidth}|p{0.16\textwidth}|p{0.48\textwidth}|}
\hline
Range of $\delta$ & Behavior of $F$ & Finite fixed points \\
\hline
$0<\delta<\dfrac{2}{2+\Delta_{\mathrm{UV}}}$
& decreasing
& one root; de Sitter attractor \\
\hline
$\dfrac{2}{2+\Delta_{\mathrm{UV}}}<\delta<1$
& single maximum
& no roots, one tangent (double) root, or two roots; the left one is a saddle point, the right one an attractor \\
\hline
$\delta>1$
& increasing
& one root; saddle point \\
\hline
\end{tabular}
\end{table*}

In the table the roots are ordered by $y$: the left one has the smaller value of $y$,
the right one the larger. In the intermediate parameter range, where
$\chi_{\mathrm{UV}}>0>\chi_{\mathrm{IR}}$, the function $F$ has a single
maximum. Its location $L_{\max}$ is determined by the conditions
\begin{equation}
\begin{array}{l}
\displaystyle \Delta_{\mathrm{loc}}(L_{\max})=\frac{2(1-\delta)}{\delta}, \\[2pt]
\displaystyle \left(\frac{L_{\max}}{L_t}\right)^\kappa=
\frac{\delta\Delta_{\mathrm{UV}}}{2(1-\delta)}-1.
\end{array}
\label{eq:Lmax}
\end{equation}
\begin{sloppypar}
In terms of the variable $x$ the function \eqref{eq:Fglobal} equals
$(x/x_*)^{\chi_{\mathrm{UV}}/\kappa}\allowbreak
\big[(1+x)/(1+x_*)\big]^{-\delta\Delta_{\mathrm{UV}}/\kappa}$, where
$x_*=(L_*/L_t)^\kappa$, so the maximum
$F_{\max}\equiv F(L_{\max})$ is known explicitly:
\end{sloppypar}
\begin{equation}
F_{\max}=\left(\frac{x_{\max}}{x_*}\right)^{\chi_{\mathrm{UV}}/\kappa}
\left(\frac{1+x_{\max}}{1+x_*}\right)^{-\delta\Delta_{\mathrm{UV}}/\kappa},
\label{eq:Fmax}
\end{equation}
where $x_{\max}=(L_{\max}/L_t)^\kappa$ is given by the second equality in
\eqref{eq:Lmax}. The condition $c_{H}^{-2}<F_{\max}$ is equivalent to
$c_{H}>c_{H,\mathrm{crit}}$ with the critical value
\begin{equation}
c_{H,\mathrm{crit}}=F_{\max}^{-1/2}.                          \label{eq:cH-crit}
\end{equation}
For $c_{H}^{-2}<F(L_{\max})$ there are two roots. The equality
$c_{H}^{-2}=F(L_{\max})$ gives a single tangent (double) root, and for
$c_{H}^{-2}>F(L_{\max})$ there are no finite roots.
The merging of the roots at $c_{H}=c_{H,\mathrm{crit}}$ corresponds to a
saddle-node bifurcation of de Sitter states; its normal
form is given below in \eqref{eq:center}. Relation \eqref{eq:Fmax}
turns the qualitative classification of Table~\ref{tab:roots} into an explicit
boundary in parameter space.
In the last
case, and for $0<\chi_{\mathrm{UV}}<2$, the solution satisfying
the global criterion has the asymptotics \eqref{eq:typeIII}.

At the tangency point $\chi_{\mathrm{dS}}=0$, so linear analysis does not determine the
stability. We introduce the deviation $\zeta=y-y_{\mathrm{dS}}$ and the coefficient
$\beta_{\mathrm{dS}}=\chi_{,y}(y_{\mathrm{dS}})$. On the invariant boundary $\Omega_X=1$,
\begin{equation}
\zeta'=\frac{\beta_{\mathrm{dS}}}{4}\zeta^2+O(\zeta^3).                                \label{eq:center}
\end{equation}
For a function of the form \eqref{eq:flow} one has $\beta_{\mathrm{dS}}<0$. Trajectories with
$\zeta>0$ approach the tangency point, while trajectories with $\zeta<0$
move away from it. On the attracting side the approach to the de Sitter
state is algebraic,
\begin{equation}
\zeta\simeq\frac{4}{|\beta_{\mathrm{dS}}|N},\quad
w_X+1\simeq-\frac{8}{3|\beta_{\mathrm{dS}}|N^2}.                           \label{eq:semistable}
\end{equation}
At the boundary value $\delta=2/(2+\Delta_{\mathrm{UV}})$ the function $F$ decreases monotonically from a finite limit as $y\to-\infty$ down to zero, while at $\delta=1$ it increases monotonically from zero up to a finite limit as $y\to+\infty$. The first of these cases corresponds to $\chi_{\mathrm{UV}}=0$, and the associated asymptotics is analyzed in \eqref{eq:typeI-boundary}. Consequently, for $c_{H}^{-2}$ smaller than the corresponding finite limit a single finite root persists; at equality the root moves to the boundary of the phase space, and above the limit there are no roots. For $\delta=1$ the finite limit is $F_{\mathrm{IR}}$, and the right boundary
can be compactified \cite{Bahamonde2018}. The root present for
$c_{H}^{-2}<F_{\mathrm{IR}}$ is a saddle point. On its right branch
$\Omega_X\to1$ and $u\to\big[c_{H}\sqrt{F_{\mathrm{IR}}}\big]^{-1}<1$.
As $t\to\infty$,
\begin{equation}
\begin{array}{l}
\displaystyle L(t)\sim\big(c_{H}\sqrt{F_{\mathrm{IR}}}-1\big)t, \\[2pt]
\displaystyle H(t)\sim
\frac{c_{H}\sqrt{F_{\mathrm{IR}}}}
{\big(c_{H}\sqrt{F_{\mathrm{IR}}}-1\big)t}, \\[2pt]
\displaystyle a(t)\propto
t^{c_{H}\sqrt{F_{\mathrm{IR}}}/
\left(c_{H}\sqrt{F_{\mathrm{IR}}}-1\right)}, \\[2pt]
\displaystyle w_X\to-\frac13-\frac{2}{3c_{H}\sqrt{F_{\mathrm{IR}}}}.
\end{array}
\nonumber
\end{equation}
This is accelerated power-law expansion with $-1<w_X<-1/3$.
Since $u$ tends to a positive constant, the global condition
\eqref{eq:global-criterion} is satisfied. At $c_{H}^{-2}=F_{\mathrm{IR}}$
the saddle point reaches the boundary of the phase space and becomes
non-hyperbolic; trajectories with finite $y$ move away from it. The left
branch for $c_{H}^{-2}<F_{\mathrm{IR}}$ and the physically admissible
trajectory for $c_{H}^{-2}\ge F_{\mathrm{IR}}$ both have the
Type III asymptotics \eqref{eq:typeIII}.

\subsection{Numerical Example}

\begin{figure}[t]
\centering
\includegraphics[width=\columnwidth]{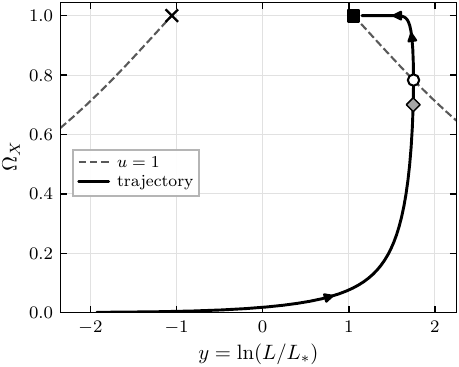}
\caption{Phase-space trajectory for the parameters
\eqref{eq:num-parameters}. The dashed line corresponds to the level set $u=1$,
that is, $\Omega_X=c_{H}^2F(y)$. The diamond marks the initial conditions,
the circle the unique crossing point, the cross the saddle
point, and the square the de Sitter attractor.}
\label{fig:phase}
\end{figure}

\begin{figure}[t]
\centering
\includegraphics[width=\columnwidth]{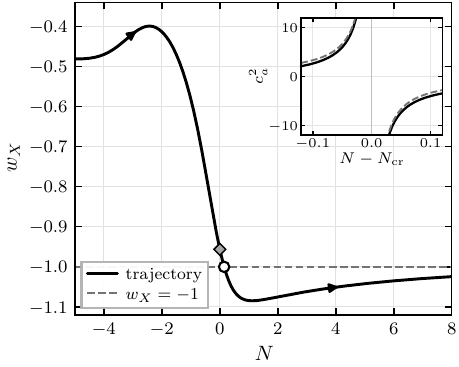}
\caption{Evolution of $w_X(N)$ along the trajectory of Fig.~\ref{fig:phase}. The diamond
marks the initial conditions and the circle the crossing point
\eqref{eq:num-crossing}; the dashed line marks the level $w_X=-1$. Inset:
the adiabatic sound speed squared $c_a^2$ near the crossing; the dashed curve is the
leading term \eqref{eq:ca-pole}
(see Sec.~\ref{sec:perturbations}).}
\label{fig:wx}
\end{figure}

The numerical example illustrates how the crossing condition and the event-horizon
boundary requirement are satisfied; no statistical fit to the data is performed here.
For comparison we first consider the model with $F=1$. For $c_{H}=1.1$
one has $u=\sqrt{\Omega_X}/c_{H}\leq c_{H}^{-1}<1$, so the
equality $u=1$ is impossible in the region $0<\Omega_X<1$. In the model with a
scale-dependent entropy we set $w_b=0$, corresponding to
pressureless matter, and choose the parameters
\begin{equation}
\begin{array}{l}
\displaystyle \delta=0.8,\quad \Delta_{\mathrm{UV}}=1,
\quad \kappa=2, \\[2pt]
\displaystyle L_t=L_*,\quad c_{H}=1.1.
\end{array}
\label{eq:num-parameters}
\end{equation}
The values are chosen for illustration: $\Delta_{\mathrm{UV}}=1$
corresponds to the Barrow limit of maximal fractal deformation, $\delta=0.8$
lies in the intermediate range of Table~\ref{tab:roots}, and for $c_{H}=1.1$
the equation $F(y)=c_{H}^{-2}$ has two roots, the right one being a
de Sitter attractor. This value excludes a crossing for $ F=1 $, so the example illustrates the effect of changing the entropy law. It does not by itself establish that a scale-dependent Barrow exponent is necessary for the crossing. Then
\begin{equation}
F(y)=(\cosh y)^{-0.4},\quad \chi(y)=-0.4\tanh y.              \label{eq:num-F}
\end{equation}
We set $N_0=0$ and take the initial conditions $y_0=1.7478$ and
$\Omega_{X0}=0.70$, which corresponds to the observed present-day dark-energy
fraction. In this subsection the subscript $0$ refers to the
initial epoch of the numerical solution; the redshift is defined
by
$1+z=a_0/a$. Then
\begin{equation}
\begin{array}{l}
\displaystyle u_0=0.94484,\quad \chi_0=-0.37645, \\[2pt]
\displaystyle w_{X0}=-0.95630.
\end{array}
\label{eq:num-initial}
\end{equation}
Numerical integration of \eqref{eq:aut1}--\eqref{eq:aut2} yields
a single crossing point (Figs.~\ref{fig:phase} and~\ref{fig:wx}):
\begin{equation}
\begin{array}{l}
\displaystyle N_{\mathrm{cr}}-N_0=0.14835,\quad \frac{a_{\mathrm{cr}}}{a_0}=1.15992, \\[2pt]
\displaystyle z_{\mathrm{cr}}=-0.13787,\quad y_{\mathrm{cr}}=1.75172, \\[2pt]
\displaystyle \Omega_{X,\mathrm{cr}}=0.78296,\quad q_{\mathrm{cr}}=-0.67444, \\[2pt]
\displaystyle w'_{X,\mathrm{cr}}=-0.25791.
\end{array}
\label{eq:num-crossing}
\end{equation}
The value $z_{\mathrm{cr}}<0$ means that the crossing lies in the future relative to the initial epoch. The negative value of $w'_{X,\mathrm{cr}}$ agrees with
the analytic relation \eqref{eq:orientation}. After the crossing the
trajectory approaches the right de Sitter root
\begin{equation}
y_{\mathrm{dS}}=1.05534,\quad \chi_{\mathrm{dS}}=-0.31355,                            \label{eq:num-attractor}
\end{equation}
which by \eqref{eq:eigenvalues} is an attractor. In the limit
$N\to\infty$ one has $u\to1$, so the integral in
\eqref{eq:global-criterion} diverges. The numerical solution therefore satisfies
the integral definition of the future event horizon.
For these parameters $ F_{\max}=1 $, so that $ c_{H,\mathrm{crit}}=1 $ according to \eqref{eq:cH-crit}.
For comparison, take
\begin{equation}
F_{\mathrm{c}}(y)=F(y_0)\exp[\chi(y_0)(y-y_0)].
\label{eq:constant-barrow-comparison}
\end{equation}
At $ \delta=0.8 $ this corresponds to a constant Barrow exponent $ \Delta_{\mathrm{c}}\simeq0.02944 $ and has the same initial $ F $ and $ \chi $. With the same $ c_H $, $ y_0 $ and $ \Omega_{X0} $, integration with $ F(y_0) $ and $ \chi(y_0) $ evaluated from \eqref{eq:num-F} without intermediate rounding gives $ N_{\mathrm{cr,c}}-N_0\simeq0.148348400 $, compared with $ N_{\mathrm{cr}}-N_0\simeq0.148347976 $. The comparison solution approaches $ y_{\mathrm{dS,c}}\simeq1.10178 $ instead of $ 1.05534 $. Thus varying the Barrow exponent is not necessary for this crossing, but changes the subsequent evolution. Imposing instead the normalization $ F_{\mathrm{c}}(0)=1 $ requires $ c_H\simeq1.23045 $ to preserve $ c_H^2F_{\mathrm{c}} $ and hence the background dynamics, not the absolute entropy.

\section{Observational Implications and Domain of Validity}

\subsection{Chevallier-Polarski-Linder Parametrization}

To compare the crossing direction with observations we return to the CPL
parametrization \eqref{eq:CPL}, which for $a_0=1$ takes the form
$w(a)=w_0+w_a(1-a)$; here $w=w_X$ and the redshift is
$z=a^{-1}-1$. In this
section the subscript $0$ refers to the present epoch and is unrelated to the
initial conditions of the numerical example.
For a local comparison we introduce the coefficient
$w_{a,\mathrm{loc}}(a)\equiv-w'_X(a)/a$, which characterizes the local
slope of $w_X(a)$ rather than a fit over a finite redshift interval. On the kinematic branch, for $w_b>-1$
and $\chi_{\mathrm{cr}}<2$, one has $w_{a,\mathrm{loc}}(a_{\mathrm{cr}})>0$.
The linear tangent approximation predicts
\begin{equation}
a_{\mathrm{cr}}^{(\mathrm{lin})}=1+\frac{1+w_0}{w_{a,\mathrm{loc},0}},\quad
z_{\mathrm{cr}}^{(\mathrm{lin})}=\big[a_{\mathrm{cr}}^{(\mathrm{lin})}\big]^{-1}-1.  \label{eq:zcpl}
\end{equation}
The exact local relation at the present epoch reads
\begin{equation}
{w_{a,\mathrm{loc},0}}=-w'_0=
\frac{\chi_{,y0}}{3}(1-u_0)^2+
\frac{2-\chi_0}{3}u'_0.                                       \label{eq:wa-map}
\end{equation}
The symbol $\chi_{,y0}$ denotes the value of
$\mathrm{d}\chi/\mathrm{d}y$ at $a=a_0$. For $u_0\simeq1$ the contribution of the
first term is quadratically small.
If the initial epoch of the numerical example is identified with the present one, \eqref{eq:num-initial} and \eqref{eq:wa-map} give $w_0\approx-0.956$ and $w_{a,\mathrm{loc},0}\approx0.330$, while the estimate \eqref{eq:zcpl} predicts $a_{\mathrm{cr}}^{(\mathrm{lin})}\approx1.13$ against the exact value $1.16$ from \eqref{eq:num-crossing}; this comparison characterizes the accuracy of the linear tangent approximation near the crossing. In the DESI DR2 analysis the combination of BAO and
CMB data indicates the region $w_0>-1$, $w_a<0$; once
supernovae are added, the statistical significance depends on the data set
\cite{DESI2025}.
The sign $w_a<0$ in the best-fit CPL models of the DESI DR2 data differs from the
sign of the local coefficient $w_{a,\mathrm{loc}}(a_{\mathrm{cr}})>0$
implied by \eqref{eq:orientation} for the kinematic branch
with $\chi_{\mathrm{cr}}<2$. The exact local statement is
\begin{equation}
\begin{array}{l}
\displaystyle \operatorname{sign}w_{a,\mathrm{loc}}(a_{\mathrm{cr}})
=\operatorname{sign}\left(2-\chi_{\mathrm{cr}}\right) \\[2pt]
\displaystyle =\operatorname{sign}\left(4-d_{S,\mathrm{cr}}\right),
\end{array}
\label{eq:discriminator}
\end{equation}
\begin{sloppypar}
so that the sign of $w_{a,\mathrm{loc}}(a_{\mathrm{cr}})$ measures the local
entropy-scaling dimension at the horizon scale, while a value $w_{a,\mathrm{loc}}(a_{\mathrm{cr}})\allowbreak<0$ would require
$d_{S,\mathrm{cr}}>4$, that is, entropy growth faster than $L^4$, which lies
outside the range \eqref{eq:corridor}. Within this range the local
coefficient is given by
\end{sloppypar}
\begin{equation}
\begin{array}{l}
\displaystyle w_{a,\mathrm{loc}}(a_{\mathrm{cr}})
=\frac{2-\chi_{\mathrm{cr}}}{2a_{\mathrm{cr}}}(1+w_b) \\[2pt]
\displaystyle \quad\times(1-\Omega_{X,\mathrm{cr}}).
\end{array}
\label{eq:wa-loc-value}
\end{equation}
The factor $(2-\chi_{\mathrm{cr}})/2$ lies in the interval $[1/2,2)$; for
$w_b=0$, $a_{\mathrm{cr}}\simeq1$ and
$\Omega_{X,\mathrm{cr}}\simeq0.7$-$0.8$ this yields
$w_{a,\mathrm{loc}}(a_{\mathrm{cr}})\approx0.1$--$0.6$. The comparison of this
local coefficient with the parameter $w_a$ obtained by fitting
\eqref{eq:CPL} to the data remains
approximate; a definitive comparison is possible only through a direct fit of the
expansion history predicted by the model.

For the late-time Universe with $w_b=0$ the same criterion
\eqref{eq:discriminator} can be written without any parametrization of $w(a)$. We introduce the jerk parameter $j=\dddot a/(aH^3)$; the
background equations give
\begin{equation}
\begin{array}{l}
\displaystyle j=1+\tfrac92\Omega_Xw_X(1+w_X)-\tfrac32\Omega_Xw_X', \\[2pt]
\displaystyle j_{\mathrm{cr}}=1-\tfrac32\Omega_{X,\mathrm{cr}}
w'_{X,\mathrm{cr}},
\end{array}
\label{eq:jerk}
\end{equation}
whence, together with \eqref{eq:crossing-inverse-map},
\begin{equation}
d_{S,\mathrm{cr}}=4+\frac{4(1-j_{\mathrm{cr}})}
{3\Omega_{X,\mathrm{cr}}(1-\Omega_{X,\mathrm{cr}})}.        \label{eq:cosmographic-test}
\end{equation}
The range \eqref{eq:corridor}, that is, $0<d_{S,\mathrm{cr}}\le3$,
then defines the local window
\begin{equation}
\begin{array}{l}
\displaystyle 1+\tfrac34\Omega_{X,\mathrm{cr}}\!\left(1-\Omega_{X,\mathrm{cr}}\right)\!
\le j_{\mathrm{cr}} \\[2pt]
\displaystyle <1+3\Omega_{X,\mathrm{cr}}\!\left(1-\Omega_{X,\mathrm{cr}}\right)\!,
\end{array}
\label{eq:jerk-window}
\end{equation}
which does not rely on the assumption that \eqref{eq:CPL} is accurate over a
finite redshift interval. For the numerical example
\eqref{eq:num-parameters}, relations \eqref{eq:jerk} and
\eqref{eq:cosmographic-test} give $j_{\mathrm{cr}}=1.30290$ and
$d_{S,\mathrm{cr}}=1.62337$, which coincides with the direct value
$2+\chi(y_{\mathrm{cr}})$ and independently confirms
\eqref{eq:crossing-inverse-map}.
A complete test of the model requires a consistent description of cosmological
perturbations.

\subsection{Necessary Thermodynamic Condition for the Phantom Regime}

On the kinematic branch the horizon radius passes through a maximum
\eqref{eq:L-series}, so after the crossing it decreases irrespective of the
sign of $2-\chi_{\mathrm{cr}}$, and the background solution must be checked for
consistency with the GSL. For an
adiabatic barotropic fluid the entropy in a comoving volume is
conserved, so the physical entropy density decreases as $a^{-3}$. The entropy $S_b$ within a sphere of radius $L$ then varies as
$S_b\propto a^{-3}L^3$, whence
\begin{equation}
\frac{S_b'}{S_b}=-3+3(1-u)=-3u,
\label{eq:Sbprime}
\end{equation}
\begin{equation}
\frac{S_h'}{S_h}=(2+\chi)(1-u).
\label{eq:Shprime}
\end{equation}
At the crossing the horizon entropy is stationary,
$S_h'=0$, whereas $S_b'=-3S_b<0$. After the crossing, for $u>1$ and
$\chi>-2$, both quantities decrease. The sum $S_h+S_b$ decreases everywhere for $u>1$, which violates the GSL.
The decisive quantity here is the sign of $1-u$: by \eqref{eq:Sbprime}--\eqref{eq:Shprime} the decrease of both entropies is equivalent to the inequality $(2+\chi)(1-u)<0$, whereas the phantom regime corresponds, by \eqref{eq:w-general}, to $(2-\chi)(1-u)<0$; these conditions are equivalent if and only if $-2<\chi<2$, and the range \eqref{eq:corridor} lies inside this interval. Outside the interval $-2<\chi<2$ the correspondence is reversed: both for $\chi>2$ and for $\chi<-2$ the decrease occurs on the quintessence side of the crossing, while the phantom phase corresponds to a growing $S_h$; on the branch $\chi=2$ the crossing occurs at $u\ne1$, and the sign of $1-u$ there requires an independent check. This is consistent with a known result: already for $F=1$ the standard definitions of temperature and entropy at the event horizon fail to ensure the first and second laws of thermodynamics, whereas at the apparent horizon both laws hold \cite{WangGongAbdalla2006,ZhouWangGongAbdalla2007}. The conclusion is sensitive to the horizon-temperature prescription: with $T=H/2\pi$ instead of $T=(2\pi L)^{-1}$, the GSL at the event horizon holds for a number of single-component models \cite{IzquierdoPavon2006}. At the crossing, where $HL=1$, the two prescriptions coincide, so the estimate below does not depend on this choice. Let $S_X$ denote an additional
dark-energy entropy.
For the total entropy to be nondecreasing, $S_X$ must satisfy the condition
\begin{equation}
S_X'\ge(2+\chi)(u-1)S_h+3uS_b,                                  \label{eq:GSL-condition}
\end{equation}
which at the crossing on the kinematic branch, where $u_{\mathrm{cr}}=1$,
reduces to $S'_{X,\mathrm{cr}}\ge3S_{b,\mathrm{cr}}$.

The standard equilibrium approach does not remedy the situation. If the dark-energy
entropy is fixed by the Gibbs relation
$T_X\,\mathrm{d}S_X=\mathrm{d}E_X+p_X\,\mathrm{d}V$ with $E_X=\rho_XV$ and
$V=\tfrac43\pi L^3$ \cite{IzquierdoPavon2006,WangGongAbdalla2006}, then
\eqref{eq:xcont} and $y'=1-u$ give
\begin{equation}
\begin{array}{l}
\displaystyle T_XS_X'=\rho_X'V+(\rho_X+p_X)V' \\[2pt]
\displaystyle =-3u(1+w_X)\rho_XV,
\end{array}
\label{eq:SXprime}
\end{equation}
and, in view of \eqref{eq:w-general}, the right-hand side equals $-u(2-\chi)(1-u)\rho_XV$.
At the crossing $w_{X,\mathrm{cr}}=-1$, so
$\rho_X+p_X=0$ and $S'_{X,\mathrm{cr}}=0$ for any finite nonzero
temperature, irrespective of its normalization. Condition \eqref{eq:GSL-condition},
however, requires $S'_{X,\mathrm{cr}}\ge3S_{b,\mathrm{cr}}>0$ there.
In the adopted decomposition $S_{\mathrm{tot}}=S_h+S_b+S_X$, the values
$S'_{h,\mathrm{cr}}=0$, $S'_{b,\mathrm{cr}}=-3S_{b,\mathrm{cr}}$ and
$S'_{X,\mathrm{cr}}=0$ give the exact equality
\begin{equation}
S'_{\mathrm{tot},\mathrm{cr}}=-3S_{b,\mathrm{cr}}<0,           \label{eq:GSL-nogo}
\end{equation}
whose negative sign is independent of $ \chi_{\mathrm{cr}} $, $ c_H $ and the temperature normalization when $ S_{b,\mathrm{cr}}>0 $. For a finite, continuous and nonzero temperature, the violation persists near the crossing. For an adiabatic barotropic fluid, this combination of additive entropy and the equilibrium Gibbs relation is incompatible with the GSL at the kinematic crossing. In nonextensive statistical mechanics the entropy composition is generally nonadditive. An additional entropy contribution or a different thermodynamic description would be required, but neither is constructed here. Energy exchange would also modify the separate conservation assumed in the background equations.

\subsection{Cosmological Perturbations}\label{sec:perturbations}

The background equations do not determine the perturbation dynamics. The simplest
test uses the barotropic equation of state
$p_X=p_X(\rho_X)$. The background adiabatic sound speed squared $c_a^2$, equal to the
derivative of the pressure with respect to the density along the solution, is
\begin{equation}
c_a^2=\frac{p_X'}{\rho_X'}=
w_X-\frac{w_X'}{3(1+w_X)}.                                     \label{eq:ca}
\end{equation}

Recall that $\tau=N-N_{\mathrm{cr}}$. Let $1+w_X$ have a zero of order $m$ at the
crossing,
\begin{equation}
1+w_X=C_m\tau^m+O(\tau^{m+1}),\quad C_m\ne0.                      \label{eq:w-zero-order}
\end{equation}
Then
\begin{equation}
\frac{w_X'}{3(1+w_X)}
={\frac{m}{3\tau}}+O(1).                                  \label{eq:ca-check}
\end{equation}
Consequently,
\begin{equation}
c_a^2=-\frac{m}{3(N-N_{\mathrm{cr}})}+O(1).                    \label{eq:ca-order}
\end{equation}
The change of sign of $1+w_X$ means that the order $m$ is odd.
For a nondegenerate crossing $m=1$, and the leading singular term is
independent of the model parameters:
\begin{equation}
c_a^2=-\frac{1}{3(N-N_{\mathrm{cr}})}+O(1),                     \label{eq:ca-pole}
\end{equation}
whereas for the degenerate cubic case \eqref{eq:cubic}, with
$m=3$, the leading term equals $-(N-N_{\mathrm{cr}})^{-1}$.
Before the crossing $c_a^2\to+\infty$, after it
$c_a^2\to-\infty$ (inset in Fig.~\ref{fig:wx}). In general $ c_a^2 $ does not coincide with the square of the physical sound speed of the perturbations. Its divergence shows that the purely barotropic adiabatic description breaks down at the crossing, but does not by itself establish physical instability. The background equations alone do not determine the stability of a more general description of perturbations. A regular description requires nonadiabatic pressure
perturbations, an effective prescription that remains regular at
$w_X=-1$, or explicit internal
degrees of freedom. This conclusion
is consistent with Vikman's no-go result for a single local scalar field
with Lagrangian density $p(\phi,X)$ \cite{Vikman2005,Hu2005}. Regular schemes of this kind are known: the parameterized post-Friedmann (PPF) framework keeps the perturbation dynamics smooth across the crossing of $w_X=-1$ \cite{Hu2008,FangHuLewis2008}.
An independent prescription of the nonadiabatic part of the pressure or of an effective dark-energy rest-frame sound speed \cite{Hu1998GDM,WellerLewis2003,BeanDore2004} itself introduces an additional assumption about the perturbation dynamics and is not equivalent to the PPF scheme \cite{Hu2008,FangHuLewis2008}. Moreover, the rest frame itself is undefined at the crossing: there the factor $\rho_X+p_X$ multiplying the dark-energy momentum density vanishes \cite{KunzSapone2006,ZhaoEtAl2005}.

\section{Conclusions}

For the dark-energy density prescription \eqref{eq:rhoX}, standard Friedmann equations, separate conservation of the components and a smooth, positive, single-valued $ F(L) $, the level $ HL=1 $ can be crossed at most once when $ w_b>-1 $ and $ 0<\Omega_X<1 $. If $ \chi<2 $ along the trajectory, every phantom-divide crossing lies on this branch and proceeds from the quintessence regime into the phantom regime. The crossing kinematics determines the local entropy-scaling dimension and the product $ c_H^2F $. The normalization of $ F $ is not determined separately. The observational restriction applies only to the local CPL coefficient. This restriction does not apply directly to a parameter fitted over a finite redshift interval.

The Barrow--Tsallis entropy function reconstructed by integrating the locally prescribed Barrow exponent has two power-law asymptotics. The condition $ \int_{N_i}^{N_f}u\,\mathrm{d}N=\infty $ is necessary and sufficient for a local horizon solution to satisfy the integral definition of the future event horizon. For this entropy function, the criterion selects de Sitter, Type III, Big Rip and accelerated power-law asymptotics under the derived parameter conditions. The numerical example satisfies this global condition. Comparison with a constant Barrow exponent yields nearly the same crossing epoch and a different de Sitter limit. A varying exponent is therefore unnecessary for the crossing in this example.

With an adiabatic barotropic fluid, additive total entropy and an equilibrium Gibbs relation for dark energy, the kinematic crossing gives $ S'_{\mathrm{tot},\mathrm{cr}}=-3S_{b,\mathrm{cr}}<0 $ for $ S_{b,\mathrm{cr}}>0 $ and finite nonzero dark-energy temperature. The divergence of $ c_a^2 $ shows that the purely barotropic adiabatic description of perturbations breaks down. Neither conclusion determines the viability of a different thermodynamic or perturbative description.

\section*{Statements and Declarations}

\paragraph{Funding.}
No funding was received for this study.

\paragraph{Competing interests.}
The sole author declares that there are no competing interests.

\paragraph{Code availability.}
All computational scripts used in this study will be provided by the author upon request.

\paragraph{Submission status.}
This manuscript is not under consideration for publication elsewhere.

\paragraph{Use of generative artificial intelligence.}
During the preparation of this manuscript, the author used ChatGPT (OpenAI) to check spelling and improve the language, style, and readability of the text. The author takes full responsibility for the content of the manuscript.

\bibliographystyle{unsrt}
\bibliography{references}

@article{CKN1999,
  author  = {Cohen, Andrew G. and Kaplan, David B. and Nelson, Ann E.},
  title   = {Effective Field Theory, Black Holes, and the Cosmological Constant},
  journal = {Phys. Rev. Lett.},
  volume  = {82},
  pages   = {4971--4974},
  year    = {1999},
  doi     = {10.1103/PhysRevLett.82.4971},
  note    = {arXiv:hep-th/9803132. \url{https://doi.org/10.1103/PhysRevLett.82.4971}}
}

@article{Li2004,
  author  = {Li, Miao},
  title   = {A Model of Holographic Dark Energy},
  journal = {Phys. Lett. B},
  volume  = {603},
  pages   = {1--5},
  year    = {2004},
  doi     = {10.1016/j.physletb.2004.10.014},
  note    = {arXiv:hep-th/0403127. \url{https://doi.org/10.1016/j.physletb.2004.10.014}}
}

@article{Wang2017,
  author  = {Wang, Shuang and Wang, Yi and Li, Miao},
  title   = {Holographic Dark Energy},
  journal = {Phys. Rep.},
  volume  = {696},
  pages   = {1--57},
  year    = {2017},
  doi     = {10.1016/j.physrep.2017.06.003},
  note    = {arXiv:1612.00345. \url{https://doi.org/10.1016/j.physrep.2017.06.003}}
}

@article{NojiriOdintsov2006,
  author = {Nojiri, Shin'ichi and Odintsov, Sergei D.},
  title = {{Unifying phantom inflation with late-time acceleration: scalar phantom-non-phantom transition model and generalized holographic dark energy}},
  journal = {Gen. Relativ. Gravit.},
  volume = {38},
  pages = {1285--1304},
  year = {2006},
  doi = {10.1007/s10714-006-0301-6},
  eprint = {hep-th/0506212},
  archivePrefix = {arXiv},
  primaryClass = {hep-th},
  note = {arXiv:hep-th/0506212. \url{https://doi.org/10.1007/s10714-006-0301-6}}
}

@article{TsallisHDE2018,
  author  = {Saridakis, Emmanuel N. and Bamba, Kazuharu and Myrzakulov, Ratbay and Anagnostopoulos, Fotios K.},
  title   = {Holographic Dark Energy through {Tsallis} Entropy},
  journal = {J. Cosmol. Astropart. Phys.},
  volume  = {2018},
  number  = {12},
  pages   = {012},
  year    = {2018},
  doi     = {10.1088/1475-7516/2018/12/012},
  note    = {arXiv:1806.01301. \url{https://doi.org/10.1088/1475-7516/2018/12/012}}
}

@article{BarrowHDE2020,
  author  = {Saridakis, Emmanuel N.},
  title   = {{Barrow} Holographic Dark Energy},
  journal = {Phys. Rev. D},
  volume  = {102},
  pages   = {123525},
  year    = {2020},
  doi     = {10.1103/PhysRevD.102.123525},
  note    = {arXiv:2005.04115. \url{https://doi.org/10.1103/PhysRevD.102.123525}}
}

@article{MamonPaliathanasisSaha2021,
  author = {Mamon, Abdulla Al and Paliathanasis, Andronikos and Saha, Subhajit},
  title = {{Dynamics of an Interacting Barrow Holographic Dark Energy Model and its Thermodynamic Implications}},
  journal = {Eur. Phys. J. Plus},
  volume = {136},
  pages = {134},
  year = {2021},
  doi = {10.1140/epjp/s13360-021-01130-7},
  eprint = {2007.16020},
  archivePrefix = {arXiv},
  primaryClass = {gr-qc},
  note = {arXiv:2007.16020. \url{https://doi.org/10.1140/epjp/s13360-021-01130-7}}
}

@article{Basilakos2025,
  author  = {Basilakos, Spyros and Lymperis, Andreas and Petronikolou, Maria and Saridakis, Emmanuel N.},
  title   = {{Barrow} Holographic Dark Energy with Varying Exponent},
  journal = {Nucl. Phys. B},
  volume  = {1015},
  pages   = {116904},
  year    = {2025},
  doi     = {10.1016/j.nuclphysb.2025.116904},
  note    = {arXiv:2312.15767. \url{https://doi.org/10.1016/j.nuclphysb.2025.116904}}
}

@article{Cimdiker2025,
  author  = {{\c C}imdiker, Ilim and D{\k a}browski, Mariusz P. and Salzano, Vincenzo},
  title   = {Generalized Nonextensive Entropy Holographic Dark Energy Models Verified by Cosmological Data},
  journal = {Eur. Phys. J. C},
  volume  = {85},
  pages   = {775},
  year    = {2025},
  doi     = {10.1140/epjc/s10052-025-14498-y},
  note    = {arXiv:2503.18230. \url{https://doi.org/10.1140/epjc/s10052-025-14498-y}}
}

@article{ZapataEtAl2026,
  author = {Zapata, Miguel A. and Garcia-Arroyo, Gabriela and Adil, Shahnawaz A. and Vazquez, J. Alberto},
  title = {{How holographic is the dark energy? A spline nodal reconstruction approach}},
  journal = {J. Cosmol. Astropart. Phys.},
  volume = {2026},
  number = {05},
  pages = {058},
  year = {2026},
  doi = {10.1088/1475-7516/2026/05/058},
  eprint = {2507.22292},
  archivePrefix = {arXiv},
  primaryClass = {astro-ph.CO},
  note = {arXiv:2507.22292. \url{https://doi.org/10.1088/1475-7516/2026/05/058}}
}

@article{KimLeeLee2012,
  author  = {Kim, Hyeong-Chan and Lee, Jae-Weon and Lee, Jungjai},
  title   = {Causality Problem in a Holographic Dark Energy Model},
  journal = {Europhys. Lett.},
  volume  = {102},
  pages   = {29001},
  year    = {2013},
  doi     = {10.1209/0295-5075/102/29001},
  note    = {arXiv:1208.3729. \url{https://doi.org/10.1209/0295-5075/102/29001}}
}

@article{Jahromi2018,
  author  = {Sayahian Jahromi, A. and Moosavi, S. A. and Moradpour, H. and Morais Gra{\c c}a, J. P. and Lobo, I. P. and Salako, I. G. and Jawad, A.},
  title   = {Generalized Entropy Formalism and a New Holographic Dark Energy Model},
  journal = {Phys. Lett. B},
  volume  = {780},
  pages   = {21--24},
  year    = {2018},
  doi     = {10.1016/j.physletb.2018.02.052},
  note    = {arXiv:1802.07722. \url{https://doi.org/10.1016/j.physletb.2018.02.052}}
}

@article{Nojiri2019,
  author  = {Nojiri, Shin'ichi and Odintsov, Sergei D. and Saridakis, Emmanuel N.},
  title   = {Modified Cosmology from Extended Entropy with Varying Exponent},
  journal = {Eur. Phys. J. C},
  volume  = {79},
  pages   = {242},
  year    = {2019},
  doi     = {10.1140/epjc/s10052-019-6740-5},
  note    = {arXiv:1903.03098. \url{https://doi.org/10.1140/epjc/s10052-019-6740-5}}
}

@article{NojiriOdintsovOikonomou2025,
  author = {Nojiri, Shin'ichi and Odintsov, Sergei D. and Oikonomou, V. K.},
  title = {{Phantom Crossing and Oscillating Dark Energy with $ F(R) $ Gravity}},
  journal = {Phys. Rev. D},
  volume = {112},
  number = {10},
  pages = {104035},
  year = {2025},
  doi = {10.1103/16yg-966k},
  eprint = {2506.21010},
  archivePrefix = {arXiv},
  primaryClass = {gr-qc},
  note = {arXiv:2506.21010. \url{https://doi.org/10.1103/16yg-966k}}
}

@article{NojiriOdintsovPaul2022,
  author  = {Nojiri, Shin'ichi and Odintsov, Sergei D. and Paul, Tanmoy},
  title   = {{Barrow} Entropic Dark Energy: A Member of Generalized Holographic Dark Energy Family},
  journal = {Phys. Lett. B},
  volume  = {825},
  pages   = {136844},
  year    = {2022},
  doi     = {10.1016/j.physletb.2021.136844},
  note    = {arXiv:2112.10159. \url{https://doi.org/10.1016/j.physletb.2021.136844}}
}

@article{NojiriOdintsovPaul2022GeneralizedEntropy,
  author = {Nojiri, Shin'ichi and Odintsov, Sergei D. and Paul, Tanmoy},
  title = {{Early and late universe holographic cosmology from a new generalized entropy}},
  journal = {Phys. Lett. B},
  volume = {831},
  pages = {137189},
  year = {2022},
  doi = {10.1016/j.physletb.2022.137189},
  eprint = {2205.08876},
  archivePrefix = {arXiv},
  primaryClass = {gr-qc},
  note = {arXiv:2205.08876. \url{https://doi.org/10.1016/j.physletb.2022.137189}}
}

@article{TsallisCirto2013,
  author  = {Tsallis, Constantino and Cirto, Leonardo J. L.},
  title   = {Black Hole Thermodynamical Entropy},
  journal = {Eur. Phys. J. C},
  volume  = {73},
  pages   = {2487},
  year    = {2013},
  doi     = {10.1140/epjc/s10052-013-2487-6},
  note    = {arXiv:1202.2154. \url{https://doi.org/10.1140/epjc/s10052-013-2487-6}}
}

@article{NojiriOdintsovFaraoni2022,
  author = {Nojiri, Shin'ichi and Odintsov, Sergei D. and Faraoni, Valerio},
  title = {{From nonextensive statistics and black hole entropy to the holographic dark universe}},
  journal = {Phys. Rev. D},
  volume = {105},
  number = {4},
  pages = {044042},
  year = {2022},
  doi = {10.1103/PhysRevD.105.044042},
  eprint = {2201.02424},
  archivePrefix = {arXiv},
  primaryClass = {gr-qc},
  note = {arXiv:2201.02424. \url{https://doi.org/10.1103/PhysRevD.105.044042}}
}

@article{Barrow2020,
  author  = {Barrow, John D.},
  title   = {The Area of a Rough Black Hole},
  journal = {Phys. Lett. B},
  volume  = {808},
  pages   = {135643},
  year    = {2020},
  doi     = {10.1016/j.physletb.2020.135643},
  note    = {arXiv:2004.09444. \url{https://doi.org/10.1016/j.physletb.2020.135643}}
}

@article{Bolotin2026,
  author       = {Bolotin, Yu. L. and Yanovsky, V. V. and Yerokhin, D. A.},
  title        = {Cosmographic Connection between Cosmological and {Planck} Scales: The {Barrow--Tsallis} Entropy},
  journal      = {J. Cosmol. Astropart. Phys.},
  volume       = {2026},
  number       = {09},
  pages        = {111},
  year         = {2026},
  doi          = {10.1088/1475-7516/2026/09/111},
  note         = {arXiv:2602.12077v3. \url{https://doi.org/10.1088/1475-7516/2026/09/111}}
}

@article{AppelquistCarazzone1975,
  author  = {Appelquist, Thomas and Carazzone, J.},
  title   = {Infrared Singularities and Massive Fields},
  journal = {Phys. Rev. D},
  volume  = {11},
  pages   = {2856--2861},
  year    = {1975},
  doi     = {10.1103/PhysRevD.11.2856},
  note = {\url{https://doi.org/10.1103/PhysRevD.11.2856}}
}

@article{Berges2002,
  author  = {Berges, J. and Tetradis, N. and Wetterich, C.},
  title   = {Non-Perturbative Renormalization Flow in Quantum Field Theory and Statistical Physics},
  journal = {Phys. Rep.},
  volume  = {363},
  pages   = {223--386},
  year    = {2002},
  doi     = {10.1016/S0370-1573(01)00098-9},
  note    = {arXiv:hep-ph/0005122. \url{https://doi.org/10.1016/S0370-1573(01)00098-9}}
}

@article{NojiriOdintsovTsujikawa2005,
  author  = {Nojiri, Shin'ichi and Odintsov, Sergei D. and Tsujikawa, Shinji},
  title   = {Properties of Singularities in the (Phantom) Dark Energy Universe},
  journal = {Phys. Rev. D},
  volume  = {71},
  pages   = {063004},
  year    = {2005},
  doi     = {10.1103/PhysRevD.71.063004},
  note    = {arXiv:hep-th/0501025. \url{https://doi.org/10.1103/PhysRevD.71.063004}}
}

@article{Bahamonde2018,
  author  = {Bahamonde, Sebastian and B{\"o}hmer, Christian G. and Carloni, Sante and Copeland, Edmund J. and Fang, Wei and Tamanini, Nicola},
  title   = {Dynamical Systems Applied to Cosmology: Dark Energy and Modified Gravity},
  journal = {Phys. Rep.},
  volume  = {775--777},
  pages   = {1--122},
  year    = {2018},
  doi     = {10.1016/j.physrep.2018.09.001},
  note    = {arXiv:1712.03107. \url{https://doi.org/10.1016/j.physrep.2018.09.001}}
}

@article{ChevallierPolarski2001,
  author  = {Chevallier, Michel and Polarski, David},
  title   = {Accelerating Universes with Scaling Dark Matter},
  journal = {Int. J. Mod. Phys. D},
  volume  = {10},
  pages   = {213--224},
  year    = {2001},
  doi     = {10.1142/S0218271801000822},
  note    = {arXiv:gr-qc/0009008. \url{https://doi.org/10.1142/S0218271801000822}}
}

@article{Linder2003,
  author  = {Linder, Eric V.},
  title   = {Exploring the Expansion History of the Universe},
  journal = {Phys. Rev. Lett.},
  volume  = {90},
  pages   = {091301},
  year    = {2003},
  doi     = {10.1103/PhysRevLett.90.091301},
  note    = {arXiv:astro-ph/0208512. \url{https://doi.org/10.1103/PhysRevLett.90.091301}}
}

@article{DESI2024,
  author  = {{DESI Collaboration}},
  title   = {{DESI} 2024 {VI}: Cosmological Constraints from the Measurements of Baryon Acoustic Oscillations},
  journal = {J. Cosmol. Astropart. Phys.},
  volume  = {2025},
  number  = {02},
  pages   = {021},
  year    = {2025},
  doi     = {10.1088/1475-7516/2025/02/021},
  note    = {arXiv:2404.03002. \url{https://doi.org/10.1088/1475-7516/2025/02/021}}
}

@article{DESI2025,
  author  = {{DESI Collaboration}},
  title   = {{DESI DR2} Results {II}: Measurements of Baryon Acoustic Oscillations and Cosmological Constraints},
  journal = {Phys. Rev. D},
  volume  = {112},
  pages   = {083515},
  year    = {2025},
  doi     = {10.1103/tr6y-kpc6},
  note    = {arXiv:2503.14738. \url{https://doi.org/10.1103/tr6y-kpc6}}
}

@article{Vikman2005,
  author  = {Vikman, Alexander},
  title   = {Can Dark Energy Evolve to the Phantom?},
  journal = {Phys. Rev. D},
  volume  = {71},
  pages   = {023515},
  year    = {2005},
  doi     = {10.1103/PhysRevD.71.023515},
  note    = {arXiv:astro-ph/0407107. \url{https://doi.org/10.1103/PhysRevD.71.023515}}
}

@article{NojiriOdintsovOikonomou2026,
  author = {Nojiri, Shin'ichi and Odintsov, Sergei D. and Oikonomou, V. K.},
  title = {{Is Phantom Divide Crossing in General Relativity Completely Impossible? Shortcomings in Canonical and Minimally Coupled Scalar Field and Possible Solutions in $ k $-Essence Models}},
  journal = {Fortschr. Phys.},
  volume = {74},
  number = {8},
  pages = {e70137},
  year = {2026},
  doi = {10.1002/prop.70137},
  eprint = {2601.21356},
  archivePrefix = {arXiv},
  primaryClass = {gr-qc},
  note = {arXiv:2601.21356. \url{https://doi.org/10.1002/prop.70137}}
}

@article{Hu2005,
  author  = {Hu, Wayne},
  title   = {Crossing the Phantom Divide: Dark Energy Internal Degrees of Freedom},
  journal = {Phys. Rev. D},
  volume  = {71},
  pages   = {047301},
  year    = {2005},
  doi     = {10.1103/PhysRevD.71.047301},
  note    = {arXiv:astro-ph/0410680. \url{https://doi.org/10.1103/PhysRevD.71.047301}}
}

@article{DESIExtended2025,
  author  = {{DESI Collaboration}},
  title   = {Extended Dark Energy Analysis Using {DESI DR2} {BAO} Measurements},
  journal = {Phys. Rev. D},
  volume  = {112},
  pages   = {083511},
  year    = {2025},
  doi     = {10.1103/w4c6-1r5j},
  note    = {arXiv:2503.14743. \url{https://doi.org/10.1103/w4c6-1r5j}}
}

@article{Turyshev2026,
  author  = {Turyshev, Slava G.},
  title   = {Dark Energy after {DESI DR2}: Observational Status, Reconstructions, and Physical Models},
  journal = {Phys. Rev. D},
  volume  = {113},
  pages   = {103540},
  year    = {2026},
  doi     = {10.1103/dqxw-yp1j},
  note    = {arXiv:2602.05368. \url{https://doi.org/10.1103/dqxw-yp1j}}
}

@article{Toomey2026,
  author  = {Toomey, Michael W. and Montefalcone, Gabriele and McDonough, Evan and Freese, Katherine},
  title   = {How Theory-Informed Priors Affect {DESI} Evidence for Evolving Dark Energy},
  journal = {Phys. Rev. D},
  volume  = {113},
  pages   = {123532},
  year    = {2026},
  doi     = {10.1103/snyr-qs56},
  note    = {arXiv:2509.13318. \url{https://doi.org/10.1103/snyr-qs56}}
}

@article{Efstathiou2025,
  author  = {Efstathiou, George},
  title   = {Evolving Dark Energy or Supernovae Systematics?},
  journal = {Mon. Not. R. Astron. Soc.},
  volume  = {538},
  pages   = {875--882},
  year    = {2025},
  doi     = {10.1093/mnras/staf301},
  note    = {arXiv:2408.07175. \url{https://doi.org/10.1093/mnras/staf301}}
}

@article{DindaMaartens2025,
  author  = {Dinda, Bikash R. and Maartens, Roy},
  title   = {Physical versus Phantom Dark Energy after {DESI}: Thawing Quintessence in a Curved Background},
  journal = {Mon. Not. R. Astron. Soc. Lett.},
  volume  = {542},
  number  = {1},
  pages   = {L31--L35},
  year    = {2025},
  doi     = {10.1093/mnrasl/slaf063},
  note = {\url{https://doi.org/10.1093/mnrasl/slaf063}}
}

@article{ShahMukherjeePal2025,
  author  = {Shah, Rahul and Mukherjee, Purba and Pal, Supratik},
  title   = {Interacting Dark Sectors in Light of {DESI DR2}},
  journal = {Mon. Not. R. Astron. Soc.},
  volume  = {542},
  pages   = {2936--2942},
  year    = {2025},
  doi     = {10.1093/mnras/staf1442},
  note    = {arXiv:2503.21652. \url{https://doi.org/10.1093/mnras/staf1442}}
}

@article{Tsujikawa2026,
  author  = {Tsujikawa, Shinji},
  title   = {Crossing the Phantom Divide in Scalar-Tensor and Vector-Tensor Theories},
  journal = {Phys. Rev. D},
  volume  = {113},
  pages   = {L041301},
  year    = {2026},
  doi     = {10.1103/y858-4swl},
  note    = {arXiv:2508.17231. \url{https://doi.org/10.1103/y858-4swl}}
}

@article{LucianoPaliathanasisSaridakis2026,
  author  = {Luciano, Giuseppe Gaetano and Paliathanasis, Andronikos and Saridakis, Emmanuel N.},
  title   = {Constraints on {Barrow} and {Tsallis} Holographic Dark Energy from {DESI DR2 BAO} Data},
  journal = {J. High Energy Astrophys.},
  volume  = {49},
  pages   = {100427},
  year    = {2026},
  doi     = {10.1016/j.jheap.2025.100427},
  note    = {arXiv:2506.03019. \url{https://doi.org/10.1016/j.jheap.2025.100427}}
}

@article{LiHDE2025,
  author  = {Li, Tian-Nuo and Li, Yun-He and Du, Guo-Hong and Wu, Peng-Ju and Feng, Lu and Zhang, Jing-Fei and Zhang, Xin},
  title   = {Revisiting Holographic Dark Energy after {DESI} 2024},
  journal = {Eur. Phys. J. C},
  volume  = {85},
  pages   = {608},
  year    = {2025},
  doi     = {10.1140/epjc/s10052-025-14279-7},
  note    = {arXiv:2411.08639. \url{https://doi.org/10.1140/epjc/s10052-025-14279-7}}
}

@article{Naik2026,
  author  = {Naik, Devaraja Mallesha},
  title   = {Confronting Holographic Dark Energy with the Latest Cosmological Data: Tensions and Model Viability},
  journal = {Mon. Not. R. Astron. Soc.},
  volume  = {547},
  number  = {3},
  pages   = {stag365},
  year    = {2026},
  doi     = {10.1093/mnras/stag365},
  note = {\url{https://doi.org/10.1093/mnras/stag365}}
}

@article{WuHDE2025,
  author       = {Wu, Peng-Ju and Li, Tian-Nuo and Du, Guo-Hong and Zhang, Xin},
  title        = {Observational Challenges to Holographic and {Ricci} Dark Energy Paradigms: Insights from {ACT} {DR6} and {DESI} {DR2}},
  journal      = {Chin. Phys. C},
  volume       = {50},
  number       = {4},
  pages        = {045105},
  year         = {2026},
  doi          = {10.1088/1674-1137/ae3be2},
  note         = {arXiv:2509.02945 [astro-ph.CO]. \url{https://doi.org/10.1088/1674-1137/ae3be2}}
}

@misc{ShlivkoPoulin2026,
  author       = {Shlivko, David and Poulin, Vivian},
  title        = {Phantom-Crossing Dark Energy and the {$\Omega_m$} Tug-of-War},
  year         = {2026},
  howpublished = {arXiv:2603.22406v1 [astro-ph.CO]}
}

@article{Mishra2026,
  author  = {Mishra, Swagat S.},
  title   = {Effective Phantom Dark Energy: What Cosmological Reconstruction Does and Does Not Imply},
  journal = {Phys. Dark Universe},
  volume  = {53},
  pages   = {102399},
  year    = {2026},
  doi     = {10.1016/j.dark.2026.102399},
  note    = {arXiv:2605.27301. \url{https://doi.org/10.1016/j.dark.2026.102399}}
}

@misc{LiEtAl2026Interactions,
  author = {Li, Tian-Nuo and Giar{\`e}, William and Du, Guo-Hong and Li, Yun-He and Di Valentino, Eleonora and Zhang, Jing-Fei and Zhang, Xin},
  title = {{Robust Preference for Dark Sector Interactions}},
  year = {2026},
  eprint = {2601.07361},
  archivePrefix = {arXiv},
  primaryClass = {astro-ph.CO},
  note = {arXiv:2601.07361}
}

@article{Chen2026,
  author  = {Chen, Ruiqi and Cline, James M. and Muralidharan, Varun and Salewicz, Benjamin},
  title   = {Quintessential Dark Energy Crossing the Phantom Divide},
  journal = {J. Cosmol. Astropart. Phys.},
  volume  = {2026},
  number  = {03},
  pages   = {044},
  year    = {2026},
  doi     = {10.1088/1475-7516/2026/03/044},
  note    = {arXiv:2508.19101. \url{https://doi.org/10.1088/1475-7516/2026/03/044}}
}

@article{Thanankullaphong2026,
  author  = {Thanankullaphong, Phusuda and Sahoo, Prasanta and Hassan Puttasiddappa, Prajwal and Roy, Nandan},
  title   = {Quintom Dark Energy: Future Attractor and Phantom Crossing in Light of {DESI DR2} Observations},
  journal = {Phys. Rev. D},
  volume  = {113},
  pages   = {084069},
  year    = {2026},
  doi     = {10.1103/4b1x-6pcx},
  note    = {arXiv:2601.02284. \url{https://doi.org/10.1103/4b1x-6pcx}}
}

@article{Quintom2010,
  author  = {Cai, Yi-Fu and Saridakis, Emmanuel N. and Setare, Mohammad R. and Xia, Jun-Qing},
  title   = {Quintom Cosmology: Theoretical Implications and Observations},
  journal = {Phys. Rep.},
  volume  = {493},
  pages   = {1--60},
  year    = {2010},
  doi     = {10.1016/j.physrep.2010.04.001},
  note    = {arXiv:0909.2776. \url{https://doi.org/10.1016/j.physrep.2010.04.001}}
}

@article{Quintom2024,
  author  = {Yang, Yuhang and Ren, Xin and Wang, Bo and Cai, Yi-Fu and Saridakis, Emmanuel N.},
  title   = {Quintom Cosmology and Modified Gravity after {DESI} 2024},
  journal = {Sci. Bull.},
  volume  = {69},
  pages   = {2698--2704},
  year    = {2024},
  doi     = {10.1016/j.scib.2024.07.029},
  note    = {arXiv:2404.19437. \url{https://doi.org/10.1016/j.scib.2024.07.029}}
}

@article{Hsu2004,
  author  = {Hsu, Stephen D. H.},
  title   = {Entropy Bounds and Dark Energy},
  journal = {Phys. Lett. B},
  volume  = {594},
  pages   = {13--16},
  year    = {2004},
  doi     = {10.1016/j.physletb.2004.05.020},
  note    = {arXiv:hep-th/0403052. \url{https://doi.org/10.1016/j.physletb.2004.05.020}}
}

@article{Bekenstein1974,
  author  = {Bekenstein, Jacob D.},
  title   = {Generalized Second Law of Thermodynamics in Black-Hole Physics},
  journal = {Phys. Rev. D},
  volume  = {9},
  pages   = {3292--3300},
  year    = {1974},
  doi     = {10.1103/PhysRevD.9.3292},
  note = {\url{https://doi.org/10.1103/PhysRevD.9.3292}}
}

@article{Davies1987,
  author  = {Davies, Paul C. W.},
  title   = {Cosmological Horizons and the Generalized Second Law of Thermodynamics},
  journal = {Class. Quantum Grav.},
  volume  = {4},
  pages   = {L225--L228},
  year    = {1987},
  doi     = {10.1088/0264-9381/4/6/006},
  note = {\url{https://doi.org/10.1088/0264-9381/4/6/006}}
}

@article{WangGongAbdalla2006,
  author  = {Wang, Bin and Gong, Yungui and Abdalla, Elcio},
  title   = {Thermodynamics of an Accelerated Expanding Universe},
  journal = {Phys. Rev. D},
  volume  = {74},
  pages   = {083520},
  year    = {2006},
  doi     = {10.1103/PhysRevD.74.083520},
  note    = {arXiv:gr-qc/0511051. \url{https://doi.org/10.1103/PhysRevD.74.083520}}
}

@article{ZhouWangGongAbdalla2007,
  author  = {Zhou, Jia and Wang, Bin and Gong, Yungui and Abdalla, Elcio},
  title   = {The Generalized Second Law of Thermodynamics in the Accelerating Universe},
  journal = {Phys. Lett. B},
  volume  = {652},
  pages   = {86--91},
  year    = {2007},
  doi     = {10.1016/j.physletb.2007.06.067},
  note    = {arXiv:0705.1264. \url{https://doi.org/10.1016/j.physletb.2007.06.067}}
}

@article{IzquierdoPavon2006,
  author  = {Izquierdo, Germ{\'a}n and Pav{\'o}n, Diego},
  title   = {Dark Energy and the Generalized Second Law},
  journal = {Phys. Lett. B},
  volume  = {633},
  pages   = {420--426},
  year    = {2006},
  doi     = {10.1016/j.physletb.2005.12.040},
  note    = {arXiv:astro-ph/0505601. \url{https://doi.org/10.1016/j.physletb.2005.12.040}}
}

@article{BarrowBBN2021,
  author  = {Barrow, John D. and Basilakos, Spyros and Saridakis, Emmanuel N.},
  title   = {Big Bang Nucleosynthesis Constraints on {Barrow} Entropy},
  journal = {Phys. Lett. B},
  volume  = {815},
  pages   = {136134},
  year    = {2021},
  doi     = {10.1016/j.physletb.2021.136134},
  note    = {arXiv:2010.00986. \url{https://doi.org/10.1016/j.physletb.2021.136134}}
}

@article{SheykhiShahbazi2025,
  author  = {Sheykhi, Ahmad and Shahbazi, Ava},
  title   = {{Barrow} Cosmology and Big-Bang Nucleosynthesis},
  journal = {Phys. Rev. D},
  volume  = {111},
  pages   = {043518},
  year    = {2025},
  doi     = {10.1103/PhysRevD.111.043518},
  note    = {arXiv:2411.06075. \url{https://doi.org/10.1103/PhysRevD.111.043518}}
}

@article{LeonEtAl2021,
  author  = {Leon, Genly and Maga{\~n}a, Juan and Hern{\'a}ndez-Almada, A. and
             Garc{\'i}a-Aspeitia, Miguel A. and Verdugo, Tom{\'a}s and Motta, V.},
  title   = {{Barrow} Entropy Cosmology: An Observational Approach with a Hint of Stability Analysis},
  journal = {J. Cosmol. Astropart. Phys.},
  volume  = {2021},
  number  = {12},
  pages   = {032},
  year    = {2021},
  doi     = {10.1088/1475-7516/2021/12/032},
  note    = {arXiv:2108.10998. \url{https://doi.org/10.1088/1475-7516/2021/12/032}}
}

@article{Hu2008,
  author  = {Hu, Wayne},
  title   = {Parametrized Post-{Friedmann} Signatures of Acceleration in the {CMB}},
  journal = {Phys. Rev. D},
  volume  = {77},
  pages   = {103524},
  year    = {2008},
  doi     = {10.1103/PhysRevD.77.103524},
  note    = {arXiv:0801.2433. \url{https://doi.org/10.1103/PhysRevD.77.103524}}
}

@article{FangHuLewis2008,
  author  = {Fang, Wenjuan and Hu, Wayne and Lewis, Antony},
  title   = {Crossing the Phantom Divide with Parameterized Post-{Friedmann} Dark Energy},
  journal = {Phys. Rev. D},
  volume  = {78},
  pages   = {087303},
  year    = {2008},
  doi     = {10.1103/PhysRevD.78.087303},
  note    = {arXiv:0808.3125. \url{https://doi.org/10.1103/PhysRevD.78.087303}}
}

@article{GuedezounmeDindaMaartens2026,
  author  = {Guedezounme, S{\^e}cloka L. and Dinda, Bikash R. and Maartens, Roy},
  title   = {Phantom Crossing or Dark Interaction?},
  journal = {J. Cosmol. Astropart. Phys.},
  volume  = {2026},
  number  = {01},
  pages   = {062},
  year    = {2026},
  doi     = {10.1088/1475-7516/2026/01/062},
  note    = {arXiv:2507.18274. \url{https://doi.org/10.1088/1475-7516/2026/01/062}}
}

@article{IbarboPerlazaEtAl2026,
  author  = {Ibarbo-Perlaza, Pedro M. and Orjuela-Quintana, J. Bayron and Palacios-C{\'o}rdoba, Jose L. and Valenzuela-Toledo, C{\'e}sar A.},
  title   = {Inconsistencies of {Tsallis} Cosmology within Horizon Thermodynamics and Holographic Scenarios},
  journal = {Phys. Rev. D},
  volume  = {113},
  pages   = {063518},
  year    = {2026},
  doi     = {10.1103/n46p-zd3j},
  note    = {arXiv:2510.00234. \url{https://doi.org/10.1103/n46p-zd3j}}
}

@article{Das2026,
  author  = {Das, Biswajit},
  title   = {Impact of the Infrared Cutoff on Structure Formation in {Tsallis} Holographic Dark Energy},
  journal = {Phys. Rev. D},
  volume  = {113},
  pages   = {123531},
  year    = {2026},
  doi     = {10.1103/52nb-z52f},
  note    = {arXiv:2604.21490. \url{https://doi.org/10.1103/52nb-z52f}}
}

@article{GuinPaulGangopadhyay2025,
  author = {Guin, Gopinath and Paul, Souvik and Gangopadhyay, Sunandan},
  title = {{Barrow holographic dark energy interacting model in the presence of radiation and matter}},
  journal = {J. Cosmol. Astropart. Phys.},
  volume = {2025},
  number = {11},
  pages = {048},
  year = {2025},
  doi = {10.1088/1475-7516/2025/11/048},
  eprint = {2507.01070},
  archivePrefix = {arXiv},
  primaryClass = {gr-qc},
  note = {arXiv:2507.01070. \url{https://doi.org/10.1088/1475-7516/2025/11/048}}
}

@article{Hu1998GDM,
  author  = {Hu, Wayne},
  title   = {Structure Formation with Generalized Dark Matter},
  journal = {Astrophys. J.},
  volume  = {506},
  pages   = {485--495},
  year    = {1998},
  doi     = {10.1086/306274},
  note    = {arXiv:astro-ph/9801234. \url{https://doi.org/10.1086/306274}}
}

@article{WellerLewis2003,
  author  = {Weller, Jochen and Lewis, Antony M.},
  title   = {Large Scale Cosmic Microwave Background Anisotropies and Dark Energy},
  journal = {Mon. Not. R. Astron. Soc.},
  volume  = {346},
  pages   = {987--993},
  year    = {2003},
  doi     = {10.1111/j.1365-2966.2003.07144.x},
  note    = {arXiv:astro-ph/0307104. \url{https://doi.org/10.1111/j.1365-2966.2003.07144.x}}
}

@article{BeanDore2004,
  author  = {Bean, Rachel and Dor{\'e}, Olivier},
  title   = {Probing Dark Energy Perturbations: The Dark Energy Equation of State and Speed of Sound as Measured by {WMAP}},
  journal = {Phys. Rev. D},
  volume  = {69},
  pages   = {083503},
  year    = {2004},
  doi     = {10.1103/PhysRevD.69.083503},
  note    = {arXiv:astro-ph/0307100. \url{https://doi.org/10.1103/PhysRevD.69.083503}}
}

@article{KunzSapone2006,
  author  = {Kunz, Martin and Sapone, Domenico},
  title   = {Crossing the Phantom Divide},
  journal = {Phys. Rev. D},
  volume  = {74},
  pages   = {123503},
  year    = {2006},
  doi     = {10.1103/PhysRevD.74.123503},
  note    = {arXiv:astro-ph/0609040. \url{https://doi.org/10.1103/PhysRevD.74.123503}}
}

@article{ZhaoEtAl2005,
  author  = {Zhao, Gong-Bo and Xia, Jun-Qing and Li, Mingzhe and Feng, Bo and Zhang, Xinmin},
  title   = {Perturbations of the Quintom Models of Dark Energy and the Effects on Observations},
  journal = {Phys. Rev. D},
  volume  = {72},
  pages   = {123515},
  year    = {2005},
  doi     = {10.1103/PhysRevD.72.123515},
  note    = {arXiv:astro-ph/0507482. \url{https://doi.org/10.1103/PhysRevD.72.123515}}
}

\end{document}